\documentclass[sigconf]{acmart}
\AtBeginDocument{%
  \providecommand\BibTeX{{%
    \normalfont B\kern-0.5em{\scshape i\kern-0.25em b}\kern-0.8em\TeX}}}

\copyrightyear{2025}
\acmYear{2025}
\setcopyright{acmlicensed}\acmConference[UIST '25]{The 38th Annual ACM Symposium on User Interface Software and Technology}{September 28-October 1, 2025}{Busan, Republic of Korea}
\acmBooktitle{The 38th Annual ACM Symposium on User Interface Software and Technology (UIST '25), September 28-October 1, 2025, Busan, Republic of Korea}
\acmDOI{10.1145/3746059.3747792}
\acmISBN{979-8-4007-2037-6/2025/09}

\usepackage{_macros}
\usepackage{multirow}
\usepackage{caption}
\usepackage{subcaption}
\usepackage{makecell}
\usepackage{color, colortbl}
\usepackage{pifont}
\usepackage{graphicx}
\newlength{\myMheight}
\usepackage{url}
\usepackage{xcolor}
\usepackage{booktabs}
\usepackage{tabularx} 
\usepackage{float}
\usepackage{caption}

\newcommand{\cmark}{\ding{51}}%

\newcommand{\name}{ATCion}
\newcommand{\TA}{TA}
\newcommand{\NA}{NA}

\begin{document}

\title[\name]{\name{}: Exploring the Design of Icon-based Visual Aids for Enhancing In-cockpit Air Traffic Control Communication}

\author{\href{https://orcid.org/0009-0008-7730-8552}{Yue Lyu}}
\email{yue.lyu@uwaterloo.ca}
\affiliation{
    \institution{University of Waterloo}
    \city{Waterloo}
    \state{ON}
    \country{Canada}
}

\author{\href{https://orcid.org/}{Xizi Wang}}
\email{l84wang@uwaterloo.ca}
\affiliation{
    \institution{University of Waterloo}
    \city{Waterloo}
    \state{ON}
    \country{Canada}
}

\author{\href{https://orcid.org/0009-0006-9406-1606}{Hanlu Ma}}
\email{hanlu.ma@connect.ust.hk}
\affiliation{
    \institution{The Hong Kong University of Science and Technology}
    \city{Hong Kong SAR}
    \state{}
    \country{China}
}

\author{\href{https://orcid.org/0000-0001-9414-9911}{Yalong Yang}}
\email{yalong.yang@gatech.edu}
\affiliation{
    \institution{Georgia Institute of Technology}
    \city{Atlanta}
    \state{Georgia}
    \country{USA}
}

\author{\href{https://orcid.org/0000-0001-5008-4319}{Jian Zhao}}
\email{jianzhao@uwaterloo.ca}
\affiliation{
    \institution{University of Waterloo}
    \city{Waterloo}
    \state{ON}
    \country{Canada}
}

\renewcommand{\shortauthors}{Lyu, et al.}

\begin{abstract}

Effective communication between pilots and air traffic control (ATC) is essential for aviation safety, but verbal exchanges over radios are prone to miscommunication, especially under high workload conditions. While cockpit-embedded visual aids offer the potential to enhance ATC communication, little is known about how to design and integrate such aids.
We present an exploratory, user-centered investigation into the design and integration of icon-based visual aids, named ATCion, to support in-cockpit ATC communication, through four phases involving 22 pilots and 1 ATC controller.
This study contributes a validated set of design principles and visual icon components for ATC messages.
In a comparative study of ATCion, text-based visual aids, and no visual aids, we found that our design improved readback accuracy and reduced memory workload, without negatively impacting flight operations; most participants preferred ATCion over text-based aids, citing their clarity, low cognitive cost, and fast interpretability. 
Further, we point to implications and opportunities for integrating icon-based aids into future multimodal ATC communication systems to improve both safety and efficiency.
\end{abstract}



\begin{teaserfigure}
    \centering
    \includegraphics[width=1.\linewidth]{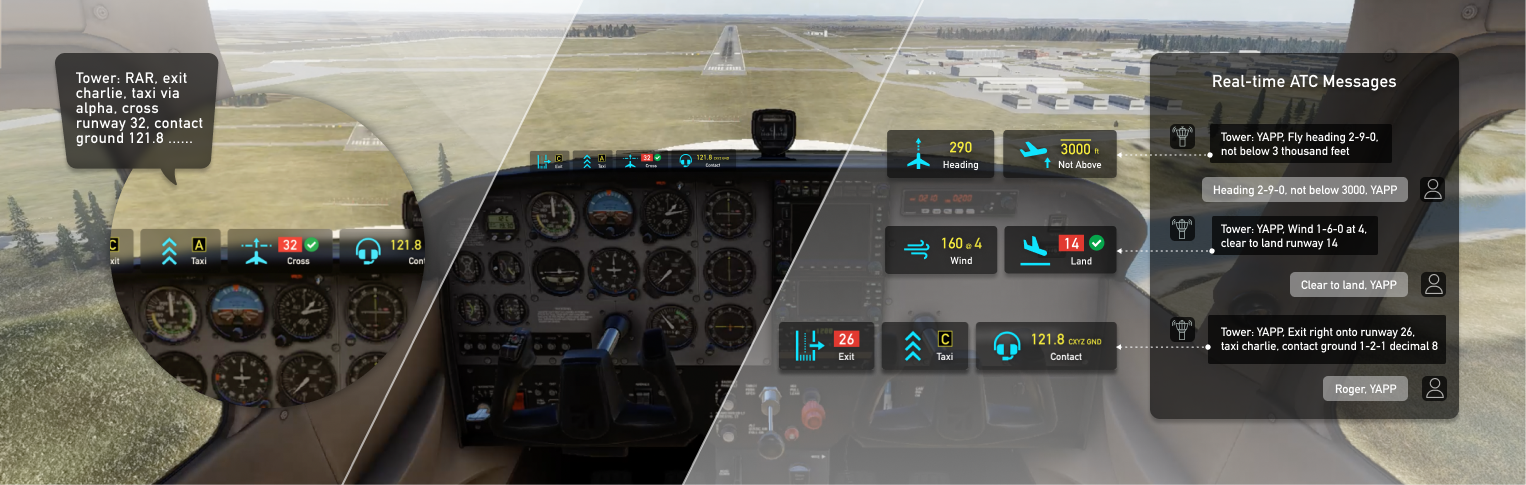}
    \vspace{-7mm}
    \caption{Through an iterative user-centered design process, we explore the design of in-cockpit icon-based visual aids to enhance pilots' traditional audio-only communication with air traffic control (ATC). Our study highlights the potential of icon-based communication to improve situational awareness, reduce the cognitive burden on pilots, and minimize communication errors in high-pressure scenarios.}
    \Description{The figure presents a panoramic view of a flight simulator cockpit featuring a real-time demonstration of icon-based visual aids for air traffic control (ATC) communication. The image is split into three sections showing different cockpit views and scenarios. Each section displays various flight instruments along with an overlay of graphical icons representing ATC instructions, such as direction to taxi, runway numbers, and ATC frequencies. The icons are designed to simplify and clarify the pilot's understanding of ATC messages, reducing cognitive load and enhancing situational awareness. Text bubbles show example ATC communications corresponding to the icons displayed, highlighting the integration of audio and visual information in high-pressure flight scenarios.}
    \label{fig:teaser}
\end{teaserfigure}
\maketitle


\section{Introduction}
Since World War II, pilots have primarily relied on audio communication with \textit{air traffic control (ATC)} to navigate the skies and ensure the safe and efficient flow of aircraft in and out of airports~\cite{FAA-ATC}.
ATC communication is vital, as pilots must continuously monitor the radio frequencies to receive crucial instructions, clearances, and real-time updates that guide their move.
The importance of this practice cannot be overstated; even the slightest miscommunication can lead to catastrophic outcomes. 
A stark reminder of this is the tragic Tenerife airport disaster~\cite{TenerifeAirportDisaster} on March 27, 1977---the deadliest accident in aviation history. 
In the dense fog of Los Rodeos Airport, a KLM Boeing 747 initiated takeoff without proper clearance, colliding with a Pan Am 747 that was still on the runway. 
The miscommunication led to the loss of 583 lives, underscoring the critical role of clear and precise ATC communication in avoiding such disasters.
Communication issues contribute to 70\% of aviation accidents and incidents~\cite{Corradini2002The}, with 73\% of runway incidents linked to misinterpreted ATC instructions and coordination failures ~\cite{Kelley1993An}.

Reducing miscommunication in ATC is essential, especially given the reliance on traditional audio communication.
ATC miscommunication can occur when messages are missed (\eg, not heard), misheard (\eg, confused with messages intended for other pilots), or misinterpreted (\eg, misunderstood clearances)~\cite{morrow1994collaboration, barshi2016misunderstandings}.
Furthermore, the complexity and length of instructions make it challenging to memorize and execute swiftly, especially under pressure, significantly increasing pilots' cognitive load and elevating the risk of miscommunication. 

Given the critical nature of the reliance on audio communication, applying the redundancy gain~\cite{lu2012redundancy} can mitigate these issues for pilots.
\textit{Visual aids} (\eg, icons and text) has been proven effective in assisting communication in general~\cite{Thibaudeau1957Visual,lipkus1999visual, Patterson1992Effects,stokes2002visual}, which has the potential to offer a valuable secondary source of ATC information. 
Specifically, leveraging multimodal perception \cite{bertelson2004psychology} by providing the information in both audio and visual representing ATC messages can reduce miscommunication~\cite{Haas2014Multimodal}.
\rev{The visual aids are designed to complement audio, especially under high workload, by reducing pilots' reliance on memory and supporting them} 
to verify details and reduce miscommunication or back-and-forth confirmations to controllers that can congest the shared radio frequency. 

Commercial systems such as Controller Pilot Data Link Communications (CPDLC)~\cite{bolczak2004controller} provide visual support by delivering ATC messages in plain text on the cockpit’s embedded computer.
However, CPDLC can only offer text-based communication for non-urgent en-route instructions.
This is not sufficient for critical flight phases and in general aviation \rev{(GA)}, where a single pilot often manages all tasks and must rely on memory or handwritten notes to capture crucial information, both of which are prone to error.

In addition to text, icons can offer an effective means of visual communication in high-stakes settings, where reducing cognitive load and ensuring rapid information processing are critical~\cite{Tao2017Are,Ferguson1971A,Chan2008Designing}.
It may also serve as quick and effective reminders, providing a promising approach to enhance real-time ATC communication. 
Compared to text alone, icons have been shown to offer advantages in recognition speed~\cite{ells1979rapid, janaka2023can}, cognitive processing efficiency~\cite{rogers1986evaluating, janaka2023can}, reducing perceived interruptions~\cite{Bailey2006On,Flynn1999Impact}, and minimizing errors~\cite{lodding1983iconic}. 
In aviation, while not particularly designed for ATC communication, icons have been used to assist remote pilots in quickly perceiving safety-critical aircraft system states~\cite{friedrich2019novel, friedrich2021human} and navigation system~\cite{Cappello2016A}. 
Despite many benefits, there are few HCI studies investigating the design of icon-based aids in ATC communication and their effectiveness, which has motivated our research.

\begin{figure*}[tb!] 
    \centering
    \includegraphics[width=1\linewidth,trim=0 10 0 10,clip]{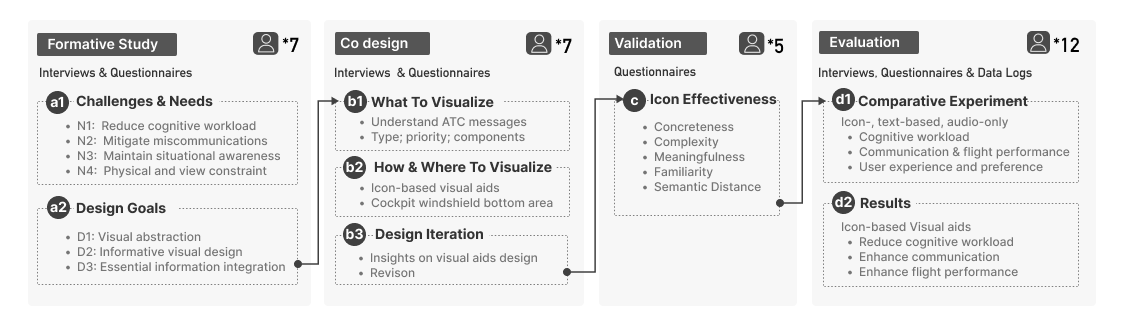}
    \vspace{-7mm}
    \caption{Our four-phase research process: a formative study identifying ATC communication needs and design goals, a co-design process developing icon-based visual aids through iterative design, a validation for design effectiveness and a user study assessing different communication conditions (audio-only, text-based, and icon-based) in VR flight simulation.}
    \Description{A structured flowchart divided into four main stages: Formative Study, Co-design, Validation, and Evaluation. Each stage is detailed with specific steps and activities. The Formative Study section, marked with a human icon and the number 7, includes interviews and questionnaires focusing on challenges and needs such as reducing cognitive workload and integrating essential information. The Co-design phase, also marked with a human icon and the number 7, involves determining what to visualize and designing iterations. Validation, indicated with a human icon and the number 5, assesses icon effectiveness through questionnaires. The final phase, Evaluation, marked with a human icon and the number 12, includes a comparative experiment and results discussion, highlighting the impact on cognitive workload, communication, and flight performance.}
    \label{fig:process}
\end{figure*}

In this study, we aim to address the gap in supporting in-cockpit ATC communication by exploring the potential of visualizations.
\rev{Specifically, our approach is designed with GA pilots in mind, where single-pilot operations are common, voice-based ATC communication remains dominant, and cockpit support tools are often limited.}
We conducted an iterative, user-centered design exploration focused on icon-based ATC visual aids (\autoref{fig:teaser}), engaging a total $N=23$ unique domain experts (\ie, pilots and ATC personnel) across \textit{four main phases} (\autoref{fig:process}).
First, to understand pilot needs, we conducted a formative user study with $N=7$ experts to identify key user needs and recurring challenges in ATC communication, and examine our initial idea of leveraging icon-based visual aids to enhance communication.
Next, we initiated a co-design process with $N=7$ experts to explore how ATC phraseology could be represented through intuitive icons, that provide pilots with clear, unambiguous depictions of ATC messages.
Together, we derived the design principles for effective iconic representation and developed \name{}, a set of visual designs that features 30 icons focusing on the most critical ATC terms augmented with essential information displays. 
Third, we validated the clarity and usability of \name{} through a follow-up study with $N=5$ experts, who rated each icon across five dimensions (\eg, concreteness, complexity, meaningfulness)~\cite{mcdougall1999measuring}, helping us refine the representations based on expert feedback.
Finally, to evaluate the effectiveness of \name{} in realistic flight contexts, we developed a functional prototype of \name{} integrated into a flight simulator. 
We conducted a within-subjects study with $N=12$ experts, comparing three communication conditions: audio-only, audio with text-based visual aids, and audio with \name{} in virtual reality (VR) flight simulations, targeting the most critical flight phases: approaching and landing.
Results from both quantitative and qualitative data showed that \name{} significantly improved pilots’ readback accuracy and reduced memory workload, particularly for longer messages. 
While flight performance remained stable across conditions, participants reported lower cognitive workload with \name{}. 
They also found the icons intuitive, effective to interpret, and helpful under pressure, especially during high-demand phases like approach and landing. 
These findings highlight the potential of icon-based aids to support situational awareness and reduce miscommunication in real-time ATC communication.

In summary, our contributions in this paper are as follows:
\begin{itemize} 
\item We conduct the first design study on icon-based aids for in-cockpit ATC communication through an iterative, user-centered process that yields practical guidelines and a range of effective artifacts, \name{}.
\item We provide rich qualitative insights into pilot communication challenges that inform future multimodal interface design to support safer and more effective ATC interactions.
\item We offer preliminary empirical evidence from a VR simulation study, with real pilots comparing the effectiveness of audio-only, text- and icon-based ATC, which lays the groundwork for future research on integrating icon-based aids into in-cockpit communication systems. 
\end{itemize} 
Through this research, we aim to offer a practical design approach to improve ATC communication, contributing to a better understanding of how visual aids can support real-time communication and enhance safety and operational efficiency in high-stakes aviation contexts.
\section{Background}
In this section, we review prior research from three perspectives: 
1) ATC miscommunication factors and current technological solutions;
2) aviation-specific visual aids;
and 3) text/icon use for improving communication across domains.

\subsection{ATC Communication Challenges}
ATC communication has evolved significantly, from ground signals to radio systems established in the 1930s~\cite{White1973Air-Ground, Sobol1984Microwave}, further advanced during World War II. 
Post-war developments included radar integration in the 1950s~\cite{Drouilhet1973The} and automated communication protocols in the 1980s~\cite{Howland1993Digital}. 
Despite these advancements, miscommunication between pilots and controllers remains a critical issue, leading to operational errors and safety risks~\cite{barshi2016misunderstandings}. 
Research shows that miscommunication occurs on average once per hour per radio frequency in high-traffic areas~\cite{Ragnarsdottir2003Language}, caused by factors such as message complexity~\cite{morrow1994collaboration}, frequency congestion~\cite{Skaltsas2013An}, language barriers~\cite{Coertze2014Aviation}, and deviations from standard protocols~\cite{barshi2016misunderstandings}.
The introduction of Controller-Pilot Data Link Communications (CPDLC) in the 2000s~\cite{bolczak2004controller} sought to reduce miscommunication by managing non-urgent communications via text, alleviating voice channel congestion~\cite{willems2006effect}. 
However, challenges remain, particularly in high-pressure, time-critical scenarios where real-time voice communication is indispensable.

As air traffic volume continues to grow~\cite{wensveen2023air}, the increasing speed and density of ATC messages impose significant cognitive demands on pilots, including the need to process, retain, and act on instructions under time constraints and in high-pressure scenarios~\cite{morrow1994collaboration}.
To alleviate these cognitive challenges, external aids such as note-taking~\cite{morrow2008designing} and message grouping~\cite{Morrow1999IMPROVING} have proven effective in supporting pilots. 
Recent advancements in automated speech recognition (ASR) systems~\cite{yang2023natural} further enhance communication by converting voice messages into text-based visual aids, facilitating the extraction of critical information and improving reliability~\cite{Lin2019Real-time}.
Building on these insights, this research investigates the potential of icon-based visual aids to further enhance real-time ATC communication. 
By designing \name{}, we aim to reduce pilots' cognitive workload, improve communication clarity, and minimize miscommunication risks in high-stakes environments.

\subsection{Visual Aids in Aviation for Safety and Efficiency}
Visual aids play a critical role in improving safety and efficiency in aviation, particularly in high-stress environments~\cite{Ferguson1971A}. 
Aids such as flight path indicators and navigational displays enable pilots to process critical information quickly, enhancing situational awareness and flight safety~\cite{wickens2021engineering}. 
Advanced systems like Synthetic Vision Systems (SVS)\cite{kramer2003synthetic} provide real-time visual representations of terrain and runways in low-visibility conditions to reduce cognitive load~\cite{Wang2020Cognitive}.
Visual alerting systems, such as the Traffic Collision Avoidance System (TCAS)\cite{Williamson1989Development, Livadas2000High-level}, deliver intuitive cues to detect threats and prevent collisions, minimizing human error in high-pressure scenarios\cite{Achachi2015TCAS}. 
Similarly, the CPDLC system~\cite{bolczak2004controller} supplements radio communication by providing text-based visual confirmations, reducing ambiguity and reinforcing comprehension.
Despite their demonstrated benefits in aviation contexts, visual aids remain underutilized in real-time ATC communication.
This presents an opportunity to explore how visual aids, particularly in dynamic flight phases like approach and landing, can enhance ATC communication efficiency and safety.

Furthermore, technologies such as Head-Up Displays (HUDs)\cite{wickens2017attentional,fischer1980cognitive} and Mixed Reality displays\cite{katins2023exploring,Rowen2019Impacts} 
enable pilots to access critical information directly within their line of sight, minimizing the need to divert attention from the external environment. 
These systems have shown significant potential in enhancing situational awareness and operational efficiency during both routine and emergency operations.
Building on these advancements, we envision a future cockpit that integrates visual displays. 
To explore this vision, we utilize VR flight simulation to study how \name{} can enhance communication efficiency by reducing miscommunication through direct visual representation.
Rather than focusing on developing new display technologies, this work aims to fill the gap by examining visual support that complements traditional verbal ATC communication.

\subsection{Enhancing Communication Efficiency with Icons and/or Text}
Icons have been extensively studied as an effective means of conveying information quickly and efficiently~\cite{hemenway1982psychological,roberts1983evaluation,whiteside1988usability,egido1988pictures,kacmar1991assessing}. 
Research shows that users can recognize pictorial representations faster and more accurately than text~\cite{zeng2009designing,hemenway1982psychological}, and icon-based systems have been found to be robust, intuitive, and expressive in human-computer communication~\cite{gittins1986icon}. Additionally, combining text labels with icons often yields optimal performance~\cite{egido1988pictures}, such as least incorrect selections in user interfaces~\cite{kacmar1991assessing}.
The proven effectiveness of icons extends across domains such as healthcare, military operations, and industrial control systems, where they enhance communication~\cite{mcdougall2000exploring}, reduce errors~\cite{wilson2012seeking}, and support faster decision-making ~\cite{galesic2009using} in time-critical contexts.
Prior studies have largely focused on cockpit displays, using icons to indicate system statuses~\cite{rogers1989icons} or navigation cues~\cite{Jones1950Pictorial}. 
For example, Friedrich \etal~\cite{friedrich2022icon} demonstrated that icons help remote pilots perceive safety-critical information more quickly, highlighting their potential to improve ATC communication.
By leveraging the benefits of combining text and icons, we developed an icon-augmented approach that abstracts ATC instructions into iconic displays while retaining essential ATC terms in text format, providing a mixed-format solution for enhancing real-time communication.

\section{Formative Study}

\begin{figure*}[tb!]
    \centering
    \includegraphics[width=1\linewidth,trim=0 10 0 10,clip]{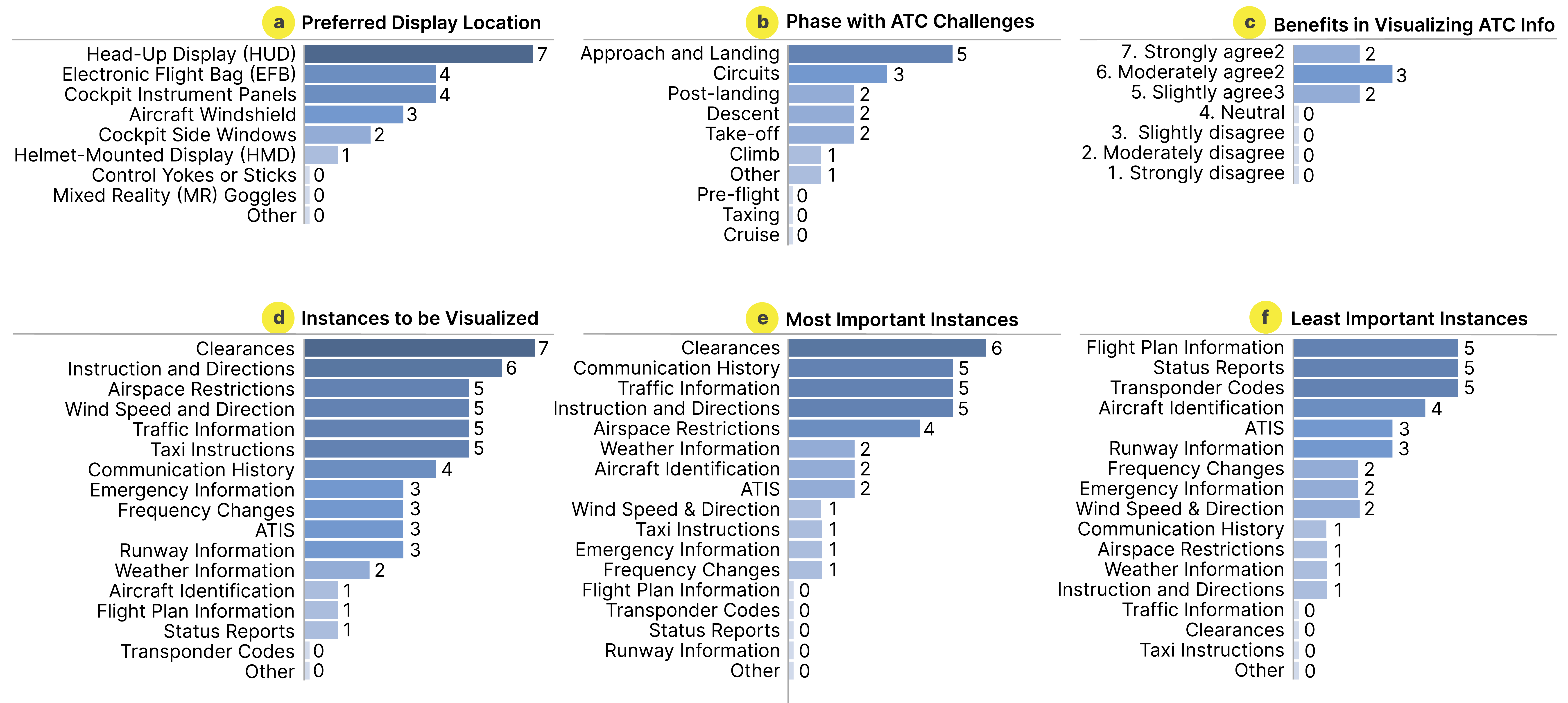}
    \vspace{-7mm}
    \caption{Questionnaire results on experts' preferences for visualization displays in the cockpit (a) and perceived benefits of different visualization options (b-d).  
    (b) indicates the phases experts found most challenging in ATC.
    (c) illustrates their agreement on the benefits of visualizing ATC information. 
    (d) shows the frequency of specific instances preferred for visualizations. 
    (e) and (f) show the rank of these instances for display in the cockpit.
    All questions allowed multiple selections, except for (c), which was single-choice.}
    \label{fig:formative_questionnaire}
    \Description{The figure comprises several panels labeled from (a) to (e), each presenting results from a questionnaire about visualization preferences and perceived benefits in cockpit displays. Panel (a) shows preferences for various visualization displays. Panel (b) highlights the phases of ATC communication found most challenging by experts, depicted through charts. Panel (c) illustrates experts' consensus on the benefits of visualizing ATC information. Panel (d) and repeated (e) display the frequency and ranking of specific instances where visualizations are preferred, shown in graphical formats. These visualizations are designed to enhance the comprehension of complex ATC communications in the cockpit environment.}
\end{figure*}

Given the specialized nature of the domain and gaps in understanding within the literature, we conducted a formative study to gain in-depth insights into the ATC communication challenges pilots face during various flight phases, their working environment in the cockpit, and early feedback on how visual aids might best support pilots in context.

This study involved a 20.5-hour semi-structured interview, supplemented by a follow-up questionnaire, with $N=7$ pilots (F1--F7; Avg. Pilot in Command (PIC)\footnote{A pilot may log PIC time when they are the sole occupant of the aircraft; are the sole manipulator of the controls of an aircraft for which the pilot is rated or has privileges; or are acting as PIC where more than one pilot is required ~\cite{FAD}.} = 72 hours), focusing on their communication experiences with ATC and interactions with existing cockpit systems. 
Additionally, two of the seven participants, who were instructor pilots, contributed valuable perspectives from their teaching experience, highlighting common trainee mistakes in ATC communication. 
A thematic analysis was conducted by two researchers in the team. 
They independently identified emergent themes from the interview transcriptions and then iteratively developed and refined codes through consensus discussions until an agreement was reached.
These findings helped identify the specific challenges and needs for ATC communication as well as the design goals for our proposed approach.

\subsection {Challenges and Needs in ATC Communication for Pilots}
Designing effective visual aids for ATC communication requires a deep understanding of pilots' natural workflows and the need to preserve their communication proficiency during routine operations. 
This is particularly important in single-pilot general aviation, where managing cognitive resources efficiently across multiple flight tasks is critical.
From the formative study, we identified the challenges in ATC communication and consolidated the following needs as well as their opinions on ATC visual aids.

\textbf{N1: Pilots need aids to reduce their cognitive workload.}
In ATC communication, pilots often rely on handwriting or memorizing critical information, which increases cognitive strain and can lead to potential errors.
F6 explained that \qt{you have this habit of taking notes when ATC gives instructions, so you don't transmit something wrong,} a point that was also echoed by F5.
Writing down information helps pilots capture key details, as F2 mentioned that \qt{when it's written down or in front of you, there's a higher chance you'll catch that.}
However, this habit of note-taking is not always feasible, as F2 pointed out that \qt{some people might not write it down, or they might not have time to write down the whole instruction. And then they miss a critical part, like holding short, especially after landing when it's quite busy.}
The fast-paced nature of operations can strain cognitive resources, leading to further challenges.
As F7 observed, \qt{[air traffic] controllers give you clearance super quickly, even before asking if you're ready to copy.} 
F6 echoed this sentiment, noting that \qt{sometimes they [ATC] speak a lot faster than they should, and that causes problems. Pilots don't understand what ATC is saying because they're too quick.}
F3 added that fatigue can exacerbate these issues: \qt{after a long day of flying or after a long flight, you just forget.}
High cognitive load can significantly impair a pilot's performance, particularly during critical phases of flight such as Approach and Landing (\autoref{fig:formative_questionnaire}).

To address these challenges, there is a clear need to reduce the pilot's cognitive load by integrating support for ATC messages, thereby reducing the risk of miscommunication. 
F4 highlighted the value of redundancy, saying \qt{pilots really like redundancy. So if there's a way to cross-check to make sure ... to just confirm.}
By providing clear aids for ATC instructions, such as long departure clearances, these cognitive challenges can be mitigated.
The aids should reduce reliance on memorization and note-taking by enabling pilots to store and recall information quickly and accurately, ensuring that they receive and understand the necessary information even in high-pressure situations.

\textbf{N2: Pilots need aids to mitigate miscommunications.}
The effectiveness of ATC communication partly depends on pilots' expectations, which can reduce cognitive workload by helping them anticipate upcoming instructions~\cite{Prinzo1993ATC/pilot}. 
However, expectation bias~\cite{Molesworth2015Miscommunication,Taylor2005Cognitive,Hamzah2018Miscommunication}, a form of miscommunication where pilots hear what they expect rather than what is actually said, can complicate communication. 
For instance, F3 noted that pilots might instinctively follow a familiar instruction, 
leading to potential issues: \qt{you expect them to say the same thing as always, like ``cross it'', and instead of holding short, you cross it. Even though you read it [hold short] back correctly, sometimes you don't fully interpret it because it's a natural reaction.}
F2, F5, and F6 also highlighted how pilots often assume clearances based on previous experiences, which can result in miscommunication if they misinterpret or overlook the actual instructions given by ATC. 
This expectation bias also affects identification. 
As F7 explained, pilots may start responding to a clearance intended for another aircraft, especially if they focus only on the first part of their call sign: \qt{they might just read back the ident, and the tower assumes it's the correct response, then moves on to the next one... We've had a couple of times where clearances go out for one plane, but another plane calls in saying copy or affirmative.}
This can cause confusion and lead to incorrect actions being taken.

To address this bias, there is a need to integrate real-time support alongside auditory messages, thereby improving clarity and reducing the risk of miscommunication. 
As F6 suggested, visual aids, such as text-based taxi instructions, can help ensure pilots accurately follow ATC guidance. 
Similarly, F4 emphasized the importance of having a way to visually confirm instructions, such as through clear visual aids for departure clearances or holding patterns.
These insights align with the questionnaire results (\autoref{fig:formative_questionnaire}), where pilots strongly agreed that visualizing ATC information, especially Clearances, Instructions, Wind and Traffic Information, would significantly aid in maintaining situational awareness and reducing cognitive load during critical flight phases. 

\textbf{N3: Pilots need aids to maintain situational awareness without distraction.}
The primary responsibility of pilots is to operate flights safely, which requires them to maintain situational awareness and minimize distractions~\cite{Endsley1999Situation}.
To achieve this, pilots must quickly and accurately perceive important information from ATC communications~\cite{Endsley2015SituationAM,Endsley1995TowardAT}.
Participants highlighted the risks of misidentifying traffic, particularly on busy days: \pqt{sometimes people will incorrectly identify with traffic and they'll miss one or two... it could be a problem...}{F5} 
F1 and F3 emphasized the potential benefits of displaying traffic information, which could assist pilots in matching visual cues with real-world traffic. 
F6 supported this, saying \qt{the pilot knows what's the weather in the area, what runways are in use, this and that, but visual aids still help.}

Effective systems should focus on reducing unnecessary communication and highlighting critical messages, thereby enhancing clarity and efficiency. 
For instance, F3 suggested that a screen showing the position of other aircraft could be particularly useful \qt{if you have a screen that shows you where the two planes [traffic] are... it's feasible if it [the screen] gives you the right direction.}
As F2 pointed out having a clear reference can save time and prevent misunderstandings, especially when a message is not clearly heard: \qt{It saves you from asking again or if you didn't catch something clearly the first time.} 
These aids should be designed to enhance, rather than hinder, a pilot's ability to perceive and respond to their environment, helping them maintain situational awareness without becoming overwhelmed.

\textbf{N4: Pilots need aids within physical and view constraints in the cockpit.}
Pilots often operate in confined spaces with limited physical movement, which significantly impacts where and how extra aids can be placed within the cockpit.
The physical layout of the cockpit is a critical factor in placing aids that do not obstruct the pilot's view or interfere with flight operation.
F3 highlighted the importance of strategic placement, noting that \qt{you need to keep your position. I wouldn't want to move around a lot... as long as you can take [the aids] with an arm's reach.}
Regarding the placement position that would not optical their workflow, F5 commented, \qt{I wouldn't say anywhere near the windows. I think it's best to just be looking outside most of the time [flying under visual flight rules]. So probably somewhere inside, maybe along the instruments.}
This suggests that visual aids should be positioned in areas where pilots naturally focus their attention, without requiring them to shift their gaze or posture significantly. 
F3 added that \qt{Side windows are fine [for display], but I'm not really looking out the side windows as often as I do with the front windows.}
This indicates that visual aids should be oriented toward the front view, where pilots spend most of their time looking.
F2 further supported this, \qt{normally on the ground, I'm looking far ahead, not straight down... [If the visual aids are] just text on the windshield, it should be fine.}
These insights underscore the importance of considering the physical limitations of the cockpit when designing visual aids. 
Ensuring that these aids are strategically placed and easily accessible is critical to enhancing pilots' efficiency and safety without disrupting their workflow.

\subsection {Design Goals}
To address these needs (N1-N4), our proposed approach focuses on improving ATC communications by reducing cognitive load, mitigating communication errors, and enhancing situational awareness through well-integrated visual displays.
In line with these priorities, we have established the following design goals to guide the creation and development of visual aids for real-time ATC communication:

\textbf{D1: Reducing cognitive workload and miscommunications with visual abstraction (N1, N2)}: 
Visual aids should effectively abstract and replicate key audio instructions in a clear, objective format to support memory and reduce cognitive strain. 
This approach aligns with research by Endsley~\cite{Endsley1999Situation} which highlights the importance of objective information presentation to prevent cognitive biases that can lead to errors (N2). 
In addition, clarity in visual design is essential to reduce cognitive load, allowing pilots to process information more efficiently (N1). 

Visual aids should effectively abstract and replicate key audio instructions in a clear, objective format to support memory and reduce cognitive strain. 
Presenting information visually helps minimize expectation bias, a known source of communication error, as highlighted in Endsley's work on objective information presentation and situation awareness~\cite{Endsley1999Situation}.
Additionally, visual clarity, achieved through concise representation and intuitive design, enables pilots to process information more efficiently under time pressure, directly addressing the challenges outlined in N1 and N2.

\textbf{D2: Enhancing situational awareness with informative visual design (N3)}: 
To avoid distraction while maintaining awareness, visual aids must strike a balance between informativeness and simplicity. This includes using readable fonts, intuitive symbols, and distinguishable colors to ensure that critical ATC information is perceptible at a glance.
Following Tufte’s principle of minimizing “chart junk”~\cite{tufte1983visual}, the design should convey essential context without unnecessary visual noise, allowing pilots to maintain focus on primary flight operations while benefiting from enhanced situational awareness.

\textbf{D3: Enabling accessibility and minimal gaze shift in limited space (N4)}:
Given the limited physical space and view angles in the cockpit, visual aids should be designed for minimal gaze shift and physical effort. 
Our formative study indicates a strong pilot preference for visualization methods such as Head-Up Displays (HUDs) that present information within the natural line of sight (\autoref{fig:formative_questionnaire}).
Thus, aids should be strategically placed near frequently monitored areas, such as the windshield or instrument panel, and designed with a compact, unobtrusive layout that aligns with pilots’ natural workflows

\section{Co-design}
To fulfill these design goals, we identified augmented reality (AR) as a promising modality for enhancing real-time ATC communication. 
AR-based visualization has the potential to display visually concise representations (D1), support through clear spatial embedding and context-rich visuals aids(D2), and remain accessible without requiring significant attention shift (D3).
Meanwhile, AR imposes constraints and opportunities including the risk of visual overload, limited spatial real estate, and the need for immediacy and glanceability all influence how visual aids should be structured, positioned, and interpreted. 
Crucially, there is currently no established framework or guideline for systematically designing and placing AR-based visualization to support in-cockpit communication. 

To address this gap, 
we adopted an iterative, user-centered co-design approach, with $N=7$ experts, of which six were experienced pilots (E1-E6, Avg. PIC = 700 hours) and one was an air traffic controller (E7), who has over 10 years of experience at an international airport.
We included the controller to gain a comprehensive understanding of ATC messages from both ATC and pilot perspectives, while maintaining our primary focus on pilots' needs.
Of the six pilots, one was from the previous formative study.
The goal of our co-design was to deeply understand the ATC messages, what information is communicated, why it is important, and how it can be conveyed effectively.
The process began with an analysis of the content and structure of ATC messages, which allowed us to identify which information is most critical to visualize.  
In addition, we explored where pilots focus across flight phases, informing the placement of the visual aids within the cockpit.
\rev{While our primary focus was on enhancing ATC communication, our design process also accounted for the broader pilot workflow. 
Guided by experts' input, we examined spatial layout, attention shifts, and timing and urgency of instructions across flight phases to ensure the design would fit naturally into multitasking contexts in GA, involving operating, navigating, and communicating.}
We then established key design principles, which together with the aforementioned design goals (D1-3) guided the development of a set of icon-based visual aids.
Throughout several design iterations, we paid attention to visual aids' readability, interpretation, and usability; any issues identified were addressed through further refinements.
Finally, we conducted a validation study where the experts rated the designed icons on five factors such as concreteness, complexity and familiarity~\cite{mcdougall1999measuring}. 
At each iteration, semi-structured interviews and questionnaires were conducted to gather domain-specific insights, ensuring that our designs were both practical and relevant to real-world aviation needs.

\begin{table*}[tb]
    \caption{Seven identified types of ATC messages, along with their priority regarding importance and where they should be displayed in the cockpit.}
\centering
\small
\vspace{-3mm}
\begin{tabular}{l|lll}
\toprule
\textbf{Message Type}             & \textbf{Description}             & \textbf{Priority}        & \textbf{Display}         \\
\midrule
Instructions             & Action-oriented messages requiring immediate attention. & High   & Within cockpit / pilot's FOV   \\
Clearances               & Authorize flight paths, altitudes, or other critical parameters.   & High   &  Within cockpit  \\
Requests                 & Inquiries for specific actions or information.   & High & Within cockpit  \\
Acknowledgments          & Confirmation of receipt and understanding of a message.    & Medium & Within cockpit   \\
General Information      & Non-time-critical information reports (\eg, weather, traffic updates).   & Medium & Within cockpit / attached to physical elements  \\
Advisories               & Alerts to potential hazards or changes in flight conditions.  & High   &  Within cockpit / attached to physical elements  \\
Emergency & Critical time-sensitive messages (\eg, go-around).    & First  & Within pilot's FOV \\    
\bottomrule
\end{tabular}
\Description{The table classifies seven types of ATC messages according to their priority and designated display locations in the cockpit. Each row represents a message type: Instructions, Clearances, Requests, Acknowledgments, General Information, Advisories, and Emergency. The descriptions clarify that Instructions, Clearances, and Requests are action-oriented or critical and have a high priority with display within the cockpit. Acknowledgments and General Information have medium priority, with display options varying slightly. Advisories also hold a high priority and may be attached to physical cockpit elements. Emergency messages, being of utmost priority, are prominently displayed within the pilot's field of view (FOV) to ensure immediate visibility and response.}

\label{tab:atc_message_priority}
\end{table*}

\subsection{What to Visualize: Understanding Air Traffic Control Messages} \label{sec:type-priority}
\subsubsection{Identifying Message Type and Priority for Visualization}
To create intuitive and informative visual abstractions of ATC messages (D1, D2), we first sought to understand their specific language. 
We reviewed standard ATC phraseology guides~\cite{vfrphraseology, ifrphraseology, icao2007radiotelephony}, and identified the types of ATC messages that pilots encounter across different flight phases via interview and questionnaire feedback from the experts. 
Together with the experts, we categorized ATC messages into seven main types: instructions, clearances, requests, acknowledgments, general information, advisories, and emergency communications (\autoref{tab:atc_message_priority}); detailed definitions are in \autoref{apx:atc-messages}. 
Through a systematic analysis of the questionnaire ratings and interviews, we gained in-depth understanding of these messages, by accessing their timeliness (\eg, how quickly the message becomes outdated), action timing (\eg, when the message should be acted upon), relevance to flight phases (\eg, during which phase the message is most likely received), and their impact on pilot workload (\eg, how busy the pilot is when receiving the message), as shown in \autoref{tab:ATC_characteristics} in \autoref{apx:atc-messages}. 
These findings provide valuable insights into how each message type should be prioritized and displayed to best support pilots in various flight phases.

Working with the experts, we further grouped the seven types of messages into three priority levels (\autoref{tab:atc_message_priority}). 
\textit{First-priority} messages (\ie, emergencies) require prominent, immediate visibility in the field of view (FOV) to ensure that pilots can respond quickly and effectively.
\textit{High-priority} messages (\ie, instructions, clearances, requests, advisories) can be displayed in fixed locations within the cockpit, allowing pilots to refer to them as needed without overwhelming their immediate workspace.
\textit{Medium-priority} messages (\ie, general information, acknowledgments) can be presented in less intrusive areas of the cockpit or as brief notifications, minimizing disruption.

\subsubsection{Identifying Critical Information for Visualization}
\label{sec:critical-info} 
To ensure the conciseness and clarity of visual aid designs (D1, D2), only the most critical aspects of ATC communication were selected for visualization to avoid overwhelming pilots operating under high cognitive workload.
Working with experts, we decomposed four types of ATC messages into their key actions or concepts and associated variables.
We started by compiling messages from the VFR (visual flight rules) and IFR (instrument flight rules) phraseology guides \cite{vfrphraseology, ifrphraseology}, and systematically deconstructing them.
Experts refined, added, or removed terms to enhance relevance and accuracy.
For example, in the message ``taxi via Alpha'', the key action is ``taxi'', while ``Alpha'' specifies the associated variable (\ie, the taxiway). 
Similarly, in ``wind 230 at 4 gusting 7'', the key concept is ``wind'', with ``230 at 4 gusting 7'' representing the wind direction and speed as essential details.
This systematic breakdown resulted in a list of key ATC terms (\autoref{tab:action}), most of which were endorsed by over half of the experts as essential for visualization.  
Terms such as ``taxi'', ``holding short'', ``cleared'', ``heading'', ``takeoff'', ``land'', and ``go-around'', were unanimously identified as critical and were prioritized in our design exploration.

\subsection{Where to Visualize: Identifying Pilots' Visual Focus Across Flight Phases} \label{sec:visual-focus}
\begin{figure}[tb]
    \centering
    \includegraphics[width=1\linewidth,trim=0 10 0 10,clip]{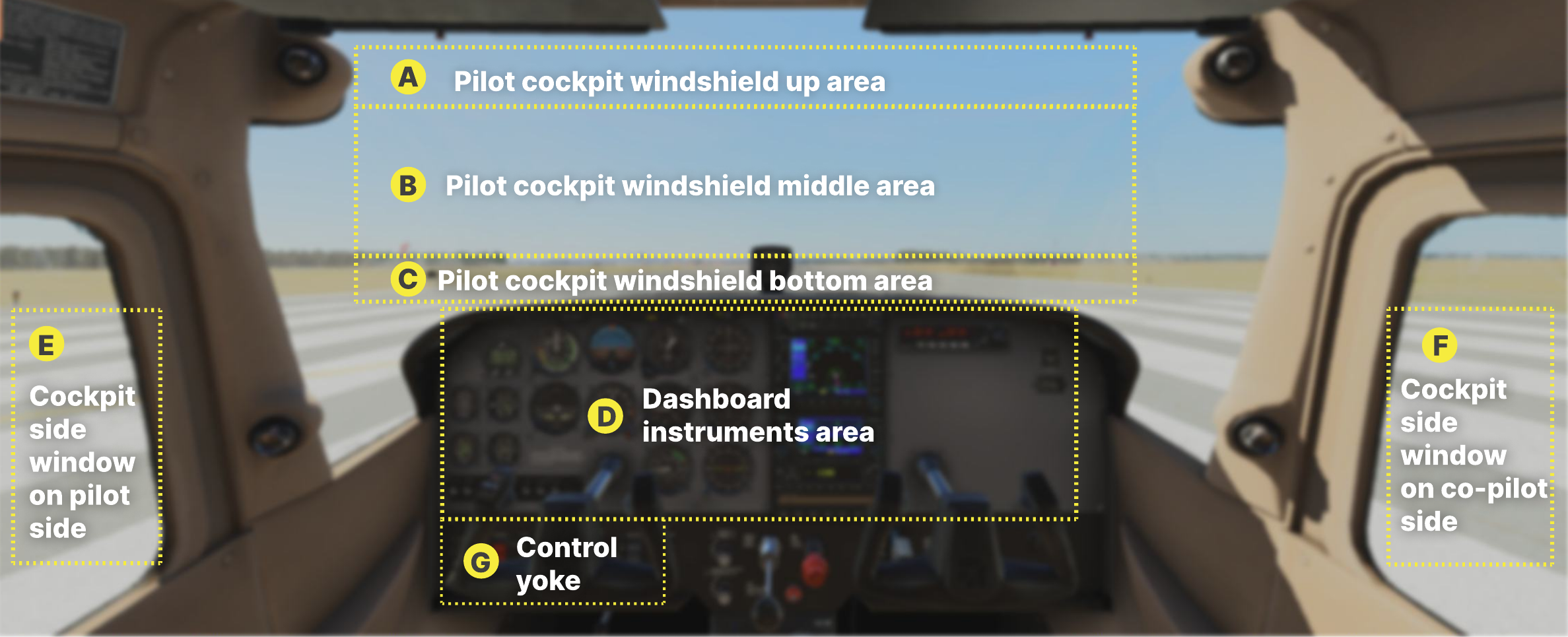}
    \vspace{-6mm}
    \caption{
       \rev{Pilot visual focus zones across the cockpit, divided into seven areas based on observed gaze patterns.  Regions include: (A) upper windshield, (B) middle windshield, (C) lower windshield, (D) dashboard, (E) side window of pilot side (F) side window on co-pilot side, and (G) control Yoke. Areas B and C serve as the primary focus zones during most flight phases.}
    }
    \label{fig:view_in_phases}
    \Description{The diagram presents the cockpit divided into seven distinct areas labeled from A to G, oriented to show pilots' visual focus during flight. Areas B (Pilot cockpit windshield bottom area) and C (Pilot cockpit windshield middle area) are emphasized as primary focus areas for pilots. Other sections include A (Pilot cockpit windshield up area), D (Dashboard instruments area), E (Control yoke), F (Cockpit side window on co-pilot side), and G (Cockpit side window on pilot side). This visualization aids in understanding how pilots distribute their visual attention across different cockpit parts during various flight phases.}
\end{figure}

Aligning visual aids with the pilot’s natural gaze patterns during flight is crucial to avoid introducing unnecessary distractions or cognitive overload (D3).
Proper placement supports smooth transitions between flight phases, ensuring pilots receive critical information at the right time. 
To determine optimal placement, we gathered questionnaire data from experts on their primary visual focus during key flight phases, including pre-flight, taxiing, takeoff, climb, cruise, descent, approach, landing, and post-landing. 
\autoref{fig:view_in_phases} illustrates key cockpit visual areas: the front windshield, divided into \rev{the upper (\autoref{fig:view_in_phases}-A), middle (\autoref{fig:view_in_phases}-B), and bottom (\autoref{fig:view_in_phases}-C) regions, the dashboard instruments (\autoref{fig:view_in_phases}-D), and the side windows (\autoref{fig:view_in_phases}-E). }
The upper windshield is used for long-range visibility during cruise, the middle windshield is critical during cruise, approach, and landing to monitor external conditions and the runway, and the bottom windshield supports taxiing, descent, and landing by maintaining focus on the immediate forward path. 
Dashboard instruments are referenced during pre-flight, descent, and approach for navigation and performance data, while side windows are essential during climbs for lateral awareness of obstacles or traffic when forward visibility is limited. 
Most pilot focus centers on the middle and bottom regions of the windshield, making these areas a priority for positioning visual aids. 
This breakdown provides clear guidelines for optimizing usability and minimizing distractions when integrating visual aids into the cockpit.

\subsection{How to Visualize: Icon-based Visual Aids Design and Validation}
Drawing from the aforementioned analyses of ATC messages and pilots' behaviors, we carried out several design iterations with our experts to develop effective icon-based visual aids. 
To probe our icon-based approach in reasonable design exploration, we selected ATC messages from the high-priority types (\autoref{sec:type-priority})---\textit{Instructions}, \textit{Clearances}, \textit{Requests}, and \textit{Advisories}, and \textit{General Information}---to design first due to their importance in real-world operations and maintain situational awareness (\autoref{fig:formative_questionnaire}). 
We also chose the nose area in the cockpit for the message display position (\autoref{fig:view_in_phases}C), because it is directly in the pilot's field of view (\autoref{sec:visual-focus}).
However, we found a notable lack of existing icons specifically tailored to representing ATC messages. 
To address this gap, we conducted a literature review~\cite{buhler_universal_2020, satcharoen_icon_2018, Pappachan2008} and identified key insights for creating clear and effective visual aids for communication.
Based on these insights and the feedback from our experts, we derived a set of principles specifically for designing icons for ATC communication.

\begin{figure*}[tb!]  
    \centering
    \includegraphics[width=0.7\linewidth,trim=0 10 0 10,clip]{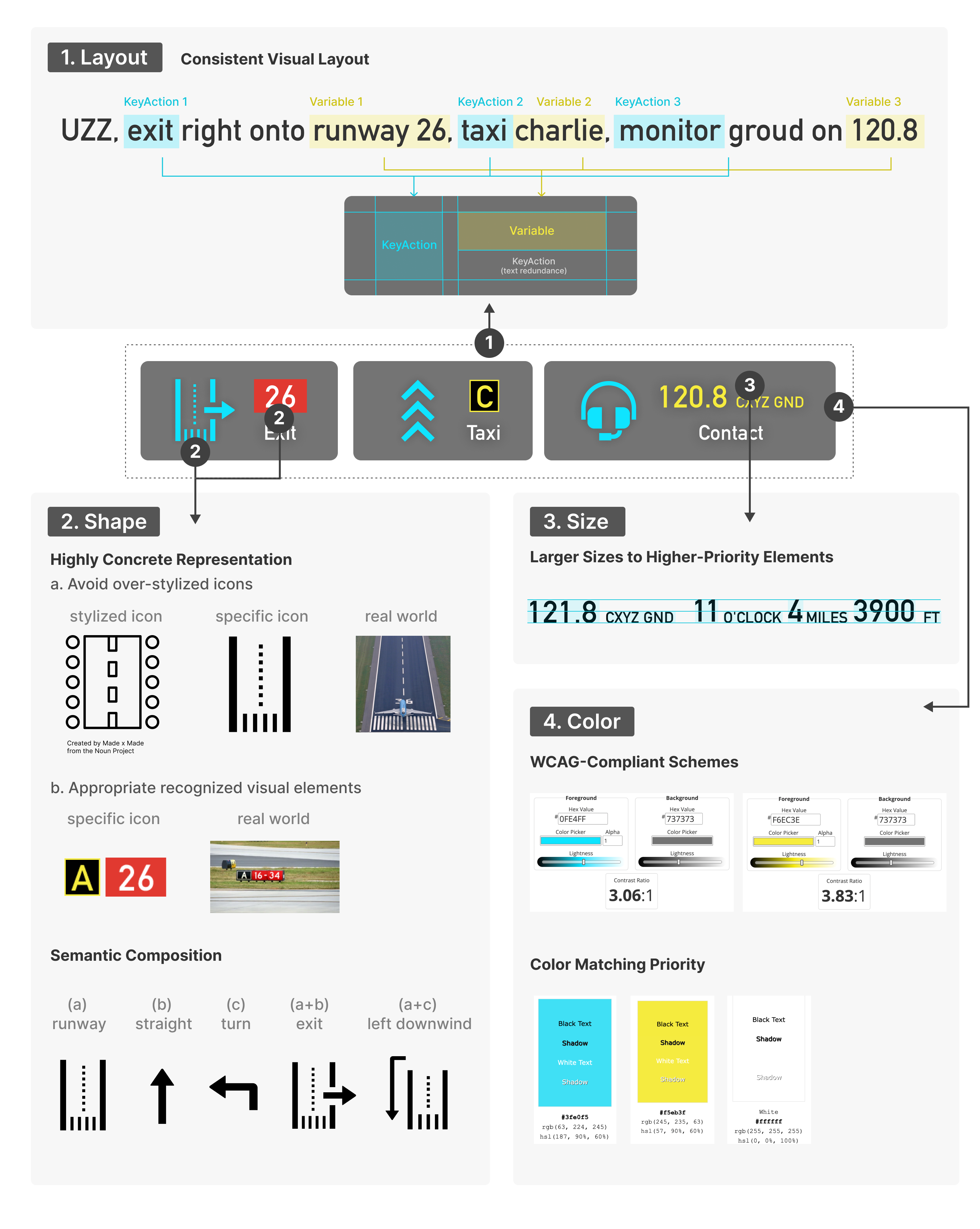}
    \vspace{-7mm}
    \caption{Overview of the design principles for the four visual channels: layout, shape, size, and color.}
    \label{fig:design_channels}
    \Description{The figure provides a comprehensive overview of four principal visual channels in cockpit display design: layout, shape, size, and color. Section 1, 'Layout,' illustrates a consistent visual layout transforming textual ATC instructions into structured visual elements. Section 2, 'Shape,' compares different levels of icon stylization from abstract to specific real-world representations. Section 3, 'Size,' highlights the use of larger sizes for higher-priority elements in a cockpit setting. Section 4, 'Color,' discusses WCAG-compliant color schemes and their application in prioritizing information based on contrast ratios. Each section is visually demarcated and includes examples and diagrams to clarify the principles discussed.}
\end{figure*} 

\begin{figure*}[tb!] 
    \centering
    \includegraphics[width=1\linewidth,trim=0 10 0 10,clip]{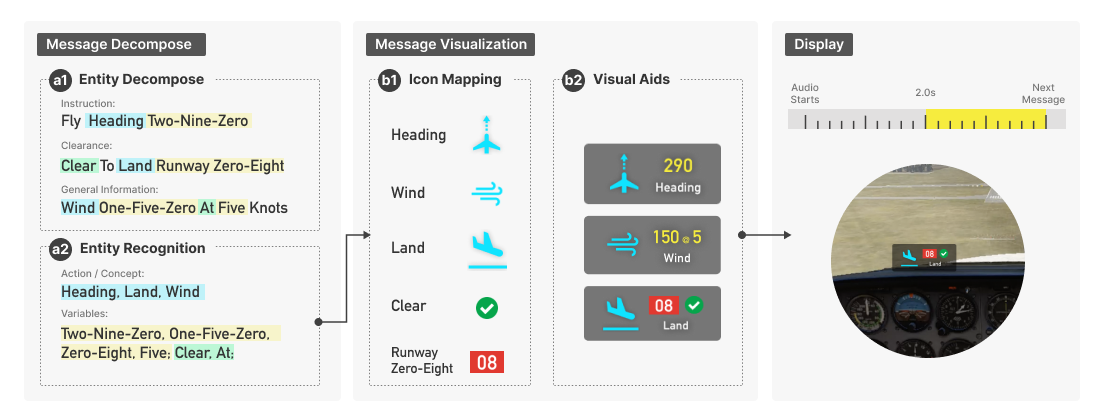}
    \vspace{-8mm}
    \caption{Process of converting an ATC message into an icon-based visual aid displayed in the cockpit.}
    \label{fig:panel_examples}
    \Description{The figure illustrates a three-step process for transforming ATC messages into visual aids within the cockpit. The first panel, 'Message Decompose', shows the breakdown of an ATC message into key entities such as 'Heading', 'Land', and 'Wind'. Each entity is then associated with specific actions or concepts. The second panel, 'Message Visualization', maps these entities to specific icons (e.g., an airplane symbol for 'Heading') and demonstrates how they are visually represented (e.g., numerical values for wind speed and direction). The final panel, 'Display', depicts how these icons are integrated into the cockpit's visual display, providing an example of a cockpit view where these icons appear in the pilot's field of view alongside an audio timeline indicating when the next message will start. This sequence highlights the efficiency of icon-based aids in enhancing pilots’ situational awareness and understanding of ATC communications.}
\end{figure*}

\subsubsection{Design Principles}
Our design principles focus on four key visual channels (\autoref{fig:design_channels}): layout, shape, size, and color. 
By optimizing these elements, we aimed to ensure the icon-based visual aids would effectively reduce pilots' workload and minimize communication errors, providing quick access to critical information during flight operations.

\textbf{Layout - Consistent Visual Layout}: 
We segmented ATC instructions into key actions/concepts and variables (\autoref{sec:critical-info}), placing corresponding visualizations in fixed positions within a grid system. A consistent layout displays the key information in a structured way, which can result in an easy learning process, reduced working memory demand, and increased efficiency~\cite{bayer_consistency_1992}. 
This is particularly beneficial when delivering long and complex instructions, because the visual aids can break them down into simple and structured panels, which are easier for pilots to recognize.

\textbf{Shape - Highly Concrete Representation}: 
Research indicates that icons require less cognitive processing to understand when they are concrete, meaning their content closely resembles real-world objects \footnote{\url{https://commons.wikimedia.org/wiki/File:McDonnell_Douglas_MD-11_KLM_-_Royal_Dutch_Airlines,_AMS_Amsterdam_(Schiphol),_Netherlands_PP1151411211.jpg}}, thereby reducing semantic distance~\cite{satcharoen_icon_2018, Pappachan2008}.
Concreteness is particularly crucial under high cognitive load, where it significantly impacts user response time and accuracy~\cite{buhler_universal_2020}. Therefore, we avoided overly stylized icons to keep semantic distance small. Additionally, we used visual elements that are widely accepted in real flight scenarios, such as the style of runway numbers \footnote{\url{https://upload.wikimedia.org/wikipedia/commons/6/69/Ben_Gurion_International_Airport_taxiway_signs.JPG}}.

\textbf{Shape - Semantic Composition}: 
As \autoref{fig:design_channels} shown, we used sub-icons combined into compound icons to convey complex and abstract concepts and key actions. For instance, icon (a) 
\wsicon{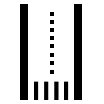}represents ``runway'', while icons (b)\wsicon{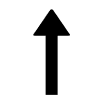}and (c) \wsicon{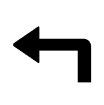} represent ``go straight'' 
and ``turn left''. When (a)\wsicon{figures/runway.png} and (b)\wsicon{figures/gostraight.png} are combined, the new icon\wsicon{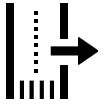} indicates ``Exit (from right of the runway)''; when (a) \wsicon{figures/runway.png}and (c) \wsicon{figures/turnleft.png}  are combined, the new icon \wsicon{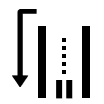}represents ``Left Downwind''. 
The repetitions of sub-icons also increase familiarity, which is an important cognitive variable that influences understanding and user performance~\cite{satcharoen_icon_2018}.

\textbf{Size - Size Matching Importance}: 
Visual elements were assigned different sizes according to their importance. For example, when visualizing lengthy variables, values were given larger font sizes, while units and suffixes, being secondary information, were assigned smaller sizes. Visually isolating key terms from other text also helps emphasize information priority and makes it more memorable~\cite{Nist1987}.

\textbf{Color - WCAG-Compliant Schemes}: 
Since pilots primarily focus on the middle to bottom areas of the front windshield, which is a dynamic area during aviation processes, good accessibility is crucial to adapt to these changes. We use a semi-transparent and blurred dark background to ensure the visual aids maintains WCAG-compliant contrast against various environments~\cite{Caldwell2008}. The color scheme ensure the readability of the visual aids and to minimize external environmental interference with ATC communication.

\textbf{Color - Color Matching Priority}: 
Visual elements were assigned colors with similar brightness but different hue and saturation according to their information priority~\cite{camgoz_effects_2004}. 
For example, key actions, having the highest priority, were set in cyan, a hue that attracts more attention. Secondary variables were set in yellow, with similar brightness and saturation but a less attention-grabbing hue. Text, as redundant information with lower priority, was set in white, a color with zero saturation.

\subsubsection{Design and Revisions}

\begin{table}[tb!]
\small
\centering
\caption{Median ratings on five dimensions of icons. 
Concreteness (Conc.): How tangible or abstract the icon appears (1: Very abstract - 5: Very concrete);
Complexity (Comp.): The level of detail in the icon (1: Very simple - 5 Very complex);
Meaningfulness (Mean.): How well the icon conveys its intended meaning (1: Not meaningful at all - 5: Very meaningful); 
Familiarity (Fam.): How common or recognizable the icon is (1:Not familiar at all - 5: Very familiar);
Semantic Distance (Sem. Dist.): The relationship between the icon and what it represents (1: Very distant - 5: Very close).
}

\vspace{-2mm}
\setlength{\tabcolsep}{4pt}
\begin{tabular}{@{}l l c c c c c@{}}
 \toprule
 \textbf{Icon} & \textbf{Meaning} & \textbf{Conc.} & \textbf{Comp.} & \textbf{Mean.} & \textbf{Fam.} & \textbf{Sem. Dist.} \\
 \midrule
  \includegraphics[width=0.04\linewidth,trim=0 2 0 0,clip]{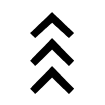} & Taxi & 4 & 2 & 4 & 2 & 4 \\ 
  \includegraphics[width=0.04\linewidth,trim=0 2 0 0,clip]{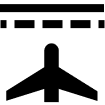} & Hold short  & 5 & 1 & 5 & 4 & 5 \\ 
  \includegraphics[width=0.04\linewidth,trim=0 2 0 0,clip]{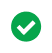} & Cleared     & 5 & 1 & 5 & 5 & 5 \\ 
  \includegraphics[width=0.04\linewidth,trim=0 2 0 0,clip]{figures/icons/clear.png} & Approved    & 5 & 1 & 5 & 5 & 5 \\ 
  \includegraphics[width=0.04\linewidth,trim=0 2 0 0,clip]{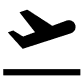} & Takeoff     & 5 & 1 & 5 & 3 & 5 \\ 
  \includegraphics[width=0.04\linewidth,trim=0 2 0 0,clip]{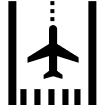} & Lineup      & 4 & 2 & 4 & 2 & 4 \\ 
  \includegraphics[width=0.04\linewidth,trim=0 2 0 0,clip]{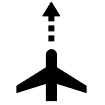} & Heading     & 5 & 2 & 4 & 5 & 5 \\ 
  \includegraphics[width=0.04\linewidth,trim=0 2 0 0,clip]{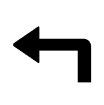}& Turn left   & 5 & 1 & 5 & 5 & 5 \\ 
  \includegraphics[width=0.04\linewidth,trim=0 2 0 0,clip]{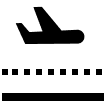} & Not below   & 4 & 2 & 4 & 4 & 4 \\ 
  \includegraphics[width=0.04\linewidth,trim=0 2 0 0,clip]{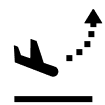} & Go-around   & 4 & 3 & 4 & 2 & 4 \\ 
  \includegraphics[width=0.04\linewidth,trim=0 2 0 0,clip]{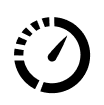} & Airspeed    & 4 & 2 & 4 & 4 & 3 \\ 
  \includegraphics[width=0.04\linewidth,trim=0 2 0 0,clip]{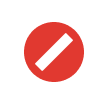} & Cancelled   & 5 & 1 & 5 & 5 & 5 \\ 
  \includegraphics[width=0.04\linewidth,trim=0 2 0 0,clip]{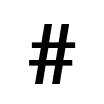} & Number      & 5 & 1 & 5 & 5 & 5 \\ 
  \includegraphics[width=0.04\linewidth,trim=0 2 0 0,clip]{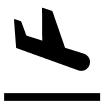}& Land        & 5 & 1 & 5 & 5 & 5 \\ 
  \includegraphics[width=0.04\linewidth,trim=0 2 0 0,clip]{figures/icons/takeoff.png}& Departure   & 5 & 1 & 5 & 5 & 5 \\ 
  \includegraphics[width=0.04\linewidth,trim=0 2 0 0,clip]{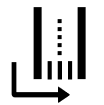}& Base        & 4 & 3 & 4 & 2 & 4 \\ 
  \includegraphics[width=0.04\linewidth,trim=0 2 0 0,clip]{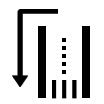}& Downwind   & 4 & 3 & 4 & 2 & 4 \\ 
  \includegraphics[width=0.04\linewidth,trim=0 2 0 0,clip]{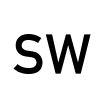}& South-West  & 5 & 1 & 5 & 5 & 5 \\ 
  \includegraphics[width=0.04\linewidth,trim=0 2 0 0,clip]{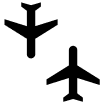}& Traffic     & 5 & 1 & 5 & 5 & 5 \\ 
  \includegraphics[width=0.04\linewidth,trim=0 2 0 0,clip]{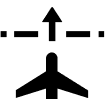}& Cross       & 4 & 2 & 5 & 4 & 4 \\ 
  \includegraphics[width=0.04\linewidth,trim=0 2 0 0,clip]{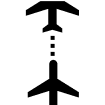}& Follow      & 5 & 1 & 5 & 4 & 4 \\ 
  \includegraphics[width=0.04\linewidth,trim=0 2 0 0,clip]{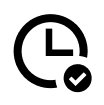}& When able   & 4 & 2 & 3 & 2 & 5 \\ 
  \includegraphics[width=0.04\linewidth,trim=0 2 0 0,clip]{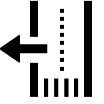}& Exit Left       & 5 & 2 & 5 & 5 & 5 \\ 
  \includegraphics[width=0.04\linewidth,trim=0 2 0 0,clip]{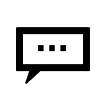}& Report      & 5 & 1 & 5 & 5 & 5 \\ 
  \includegraphics[width=0.04\linewidth,trim=0 2 0 0,clip]{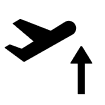}& Climb       & 5 & 1 & 5 & 5 & 5 \\ 
  \includegraphics[width=0.04\linewidth,trim=0 2 0 0,clip]{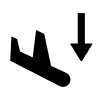}& Descend     & 5 & 1 & 5 & 5 & 5 \\ 
  \includegraphics[width=0.04\linewidth,trim=0 2 0 0,clip]{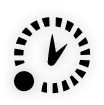}& Altimeter   & 4 & 2 & 4 & 4 & 4 \\ 
  \includegraphics[width=0.04\linewidth,trim=0 2 0 0,clip]{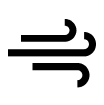}& Wind        & 5 & 1 & 5 & 5 & 5 \\ 
   \includegraphics[width=0.04\linewidth,trim=0 2 0 0,clip]{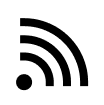}& Frequency     & 5 & 1 & 4 & 4 & 4 \\ 
  \includegraphics[width=0.04\linewidth,trim=0 2 0 0,clip]{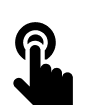}& Squawk      & 5 & 2 & 5 & 4 & 5 \\
  \includegraphics[width=0.04\linewidth,trim=0 2 0 0,clip]{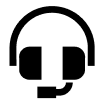}& Contact     & 5 & 1 & 5 & 4 & 5 \\ 
\bottomrule
\Description{The table presents median ratings for various ATC message icons across five dimensions: Concreteness, Complexity, Meaningfulness, Familiarity, and Semantic Distance. Each row lists an icon alongside its corresponding meaning, and numerical ratings from 1 to 5 for each dimension. Icons include visual representations for commands like Taxi, Hold Short, Cleared, and others, with most receiving high scores on Concreteness, Meaningfulness, and Semantic Distance, indicating they are tangible, meaningful, and closely related to what they represent. Complexity and Familiarity ratings vary, with some icons being simpler and less recognized than others. This table helps quantify how effectively each icon conveys its intended aviation communication function.}
\label{tab:icon-ratings}
\end{tabular}
\end{table}

Based on the principles, we developed panels (\autoref{fig:panel_examples}b2) representing common ATC messages and a set of icons (\autoref{tab:icon-ratings}) representing the key terms commonly used in ATC communications, and iteratively refined our design based on experts' feedback.

For our initial versions 
, while the experts appreciated the overall design (e.g., font, size, color), they suggested reducing the margins for a more compact layout. 
They also recommended grouping related panels or combining them into a single panel. 
They suggested incorporating directional cues (\eg, 12 o'clock or left/right) with arrows in the visual aids. Additionally, they advised including ICAO airport codes and abbreviations (\eg, GND (ground), TWR (tower)) to indicate the contact.
Moreover, the experts provided feedback on the clarity and appropriateness of each icon. 
Several key insights were identified during the process, which guided our revisions. 
For example, incorporating familiar symbols helps reduce the cognitive load on pilots and improves the speed and accuracy with which they interpret visual aids. 
Moreover, keeping complexity low while maintaining high concreteness is crucial and also a trade-off for ensuring that icons are be quickly understood without ambiguity.
The importance of designing icons needs to be conceptually aligned with their real-world counterparts. 
For icons like ``Crosswind'', introducing more specific visual metaphors that pilots encounter in their flight operations will likely improve recognition and usability. 

Therefore, over the course, we made many revisions to our visual aid designs.
We condensed the panels by reducing margins and removing the non-important information, added arrows for direction (\eg, arrow on the exit direction), grouped related info (\eg, traffic information), and included ICAO codes in frequency switching instructions.
For icon designs, in the case of the `Go-Around' icon, E1 and E6 recommended following the curved path found in Jeppesen's flight charts and airway manual~\cite{jeppesen_charts_airway_manual}
that pilots commonly use, leveraging this familiarity to improve recognition.
Similarly, the ``Hold Short'' icon was refined by adding dash lines that mapped to the real-world level of abstraction pilots are accustomed to, increasing its clarity.
In addition, we adjusted the perspective of some icons (\eg, lineup, exit) to match the pilot's view, allowing for quicker recognition during flight. 

These refinements led to our final design of the icon-based visual aids for ATC communication (\autoref{fig:panel_examples} and \autoref{tab:icon-ratings}). 
Each ATC message is broken down into its action/concept and associated variables (\autoref{fig:panel_examples}a1,2), displayed on the left and right sections of a unified-looking panel, respectively (\autoref{fig:panel_examples}b1,2). 
Actions or concepts are represented by icons for easier interpretation, and the associated variables are conveyed using appropriate symbols.
Given aviation's rigorous safety standards, we placed the text of the action/concept below the associated variables as redundant verification to ensure accurate interpretation even during high-stress operations.

\subsubsection{Initial Icon Design Validation}
To validate the effectiveness of the icons, a key element in our designed visual aids, we asked the experts to fill in a questionnaire on assessing the designed icons across five dimensions: Concreteness, Complexity, Meaningfulness, Familiarity, and Semantic Distance, on a 5-point Likert Scale~\cite{mcdougall1999measuring}.
Five out of the six pilots in our co-design process responded.
The results (\autoref{tab:icon-ratings}) show that most of the icon designs such as ``Hold Short'', ``Cleared'', ``Cancelled'', ``Number'', ``Land'', ``Departure'', ``South-West (directions)'', ``Traffic'', ``Climb'', ``Descend'', and ``Wind'' consistently received high scores (mostly 5's) across all dimensions, particularly in concreteness, and familiarity. 
These icons were highly rated for their clarity, simplicity, and meaningfulness, with a low semantic distance between the icon and its intended meaning, making them easily recognizable and useful in operational settings.
However, some icons like ``When Able'' scored lower in meaningfulness due to the abstract nature of the concept it represent. 
Additionally, icons like ``Airspeed'', which were designed based on the airspeed instrument, received lower familiarity ratings, particularly from commercial airline pilots who are less familiar with general aviation instrumentation. 
This suggests that more common visual metaphors may be necessary to improve familiarity and reduce semantic distance.
Overall, the validation results indicate that icons are concrete, simple, meaningful and familiar to pilots, confirming their suitability for operational use with a few icons (\eg, ``Base'', ``Crosswind'') that may benefit from further simplification to enhance recognition in a dynamic environment.



\section{User Study}
To investigate the effectiveness of visual aids for in-cockpit ATC communication, we developed a prototype system, \name{}, which integrates our designed icon-based visual representations tailored to pilots’ needs.
Rather than aiming to optimize existing systems, our goal is to understand how design can meaningfully support in high-stakes, real-time communication scenarios.
To evaluate the impact of \name{}, we conducted a within-subject study in a simulated cockpit environment, comparing three conditions: \name{} (icon-based visual aids), text-based visual aids, and a baseline with no visual aids.

\subsection{Participants}
We recruited $N =12$ pilots (2 females and 10 males) with diverse flight experience, ranging from student pilots with zero Pilot-In-Command (PIC) hours to experienced aviators with up to 1300 hours (\rev{\mean{210.8} h, \sd{359.5} h, \md{87.5} h}).
\rev{Among them,
2 participants had fewer than 40 PIC hours\footnote{pre-private pilot or early-stage student pilots. 40-hour minimum for private pilot licensing ~\cite{faa_private}.},
7 had between 40 - 250 hours\footnote{licensed private pilots or those progressing toward commercial certification.  250-hour requirement for commercial pilot certification ~\cite{faa_commercial}.},
and 3 had more than 250 hours\footnote{highly experienced pilots.}, thresholds that correspond to key certification milestones \cite{faa_private, faa_commercial}.
All participants had experience operating under VFR conditions with the Cessna 152 aircraft, which aligns with the simulated scenario used in our study.}
Participants were recruited through a combination of targeted outreach and open invitations. 
We approached flight schools, pilots training programs, and professional networks, ensuring representation from both general aviation and commercial aviation sectors, ensuring a broad spectrum of experience and perspectives. 
Recruitment materials were distributed via email, posters, social media, and a website, providing an overview of the study and eligibility requirements.
None of the recruited participants were from previous phases of our study.
The study has received clearance from our institutional research ethics office.

\subsection{Study Software and Data}
The study was conducted in person in a controlled environment equipped with a high-fidelity VR flight simulation, following the practice of previous study~\cite{oberhauser2017virtual}, using the X-Plane 12 flight simulation software~\cite{xplane} and an Oculus Quest 2 headset.
\rev{We adopted this setup to enable safe and controlled environment while mitigating the logistical and ethical challenges of field studies, which is a common practice in early-stage research for safety-critical domains like aviation.}
To closely replicate real-world cockpit conditions, the VR flight simulation incorporated live ATC audio recordings, providing participants with ATC messages that are typical as actual flight operations.
We collected real-world ATC audios from a local airport on public radio channels, and edited and remixed them based on our study setup. 
It would be impractical to include all ATC messages and all flight phases in a single study session, which results in a very long experiment. 
Thus, we designed the flight simulation as high-pressure scenarios by focusing on the approach and landing phases, where cognitive workload peaks and communication with ATC is critical and challenging (\autoref{fig:formative_questionnaire}).
We crafted three ATC audio scenarios with a similar level of complexity, each including background ATC communications between the tower and other pilots as well as specific audio instructions directed at the participants.
Each audio featured five messages, four short and one long, which are commonly used during these phases.

\begin{figure}[tb!]
    \centering
    \includegraphics[width=0.32\linewidth,trim=0 10 0 10,clip]{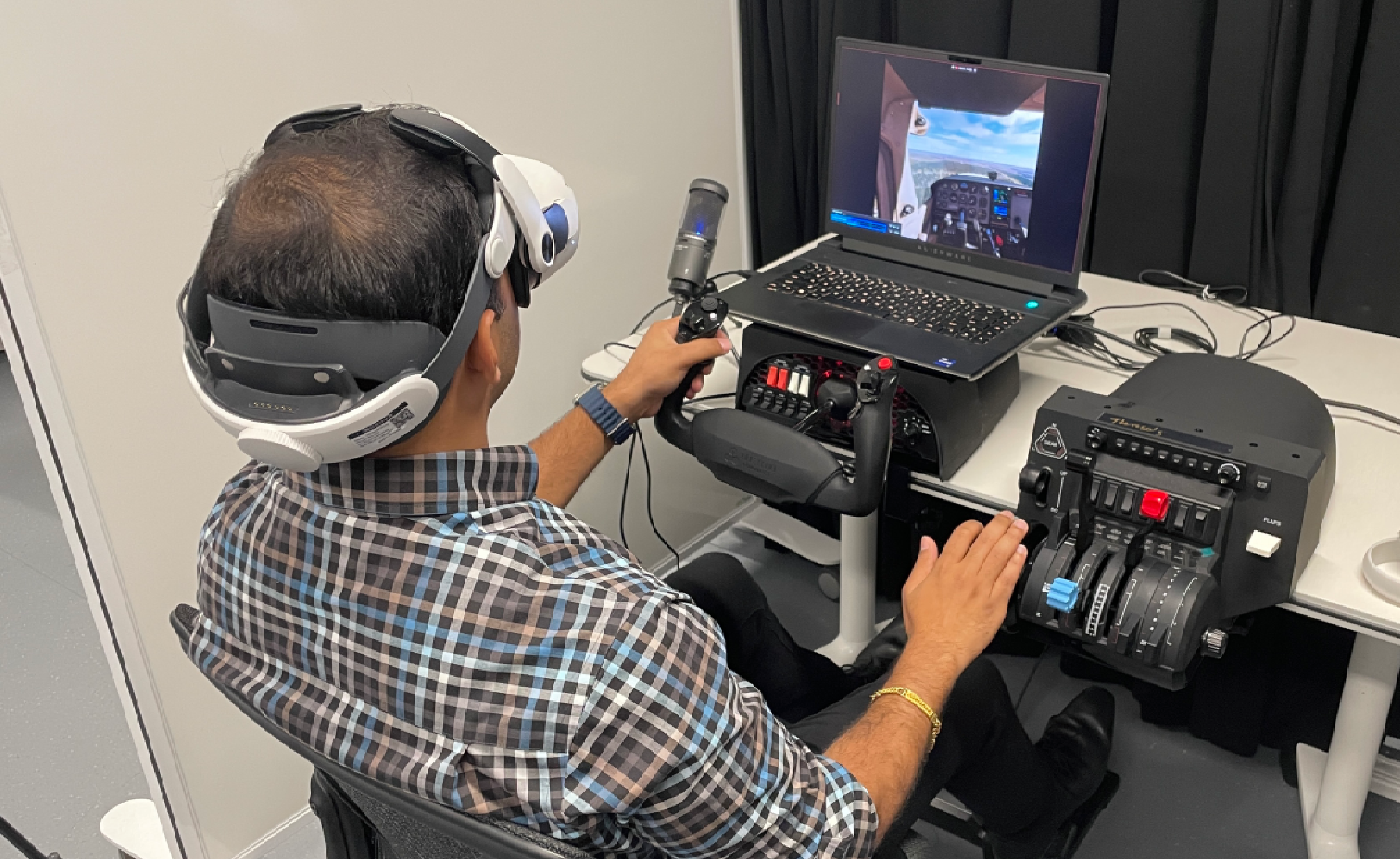}
    \includegraphics[width=0.32\linewidth,trim=0 10 0 10,clip]{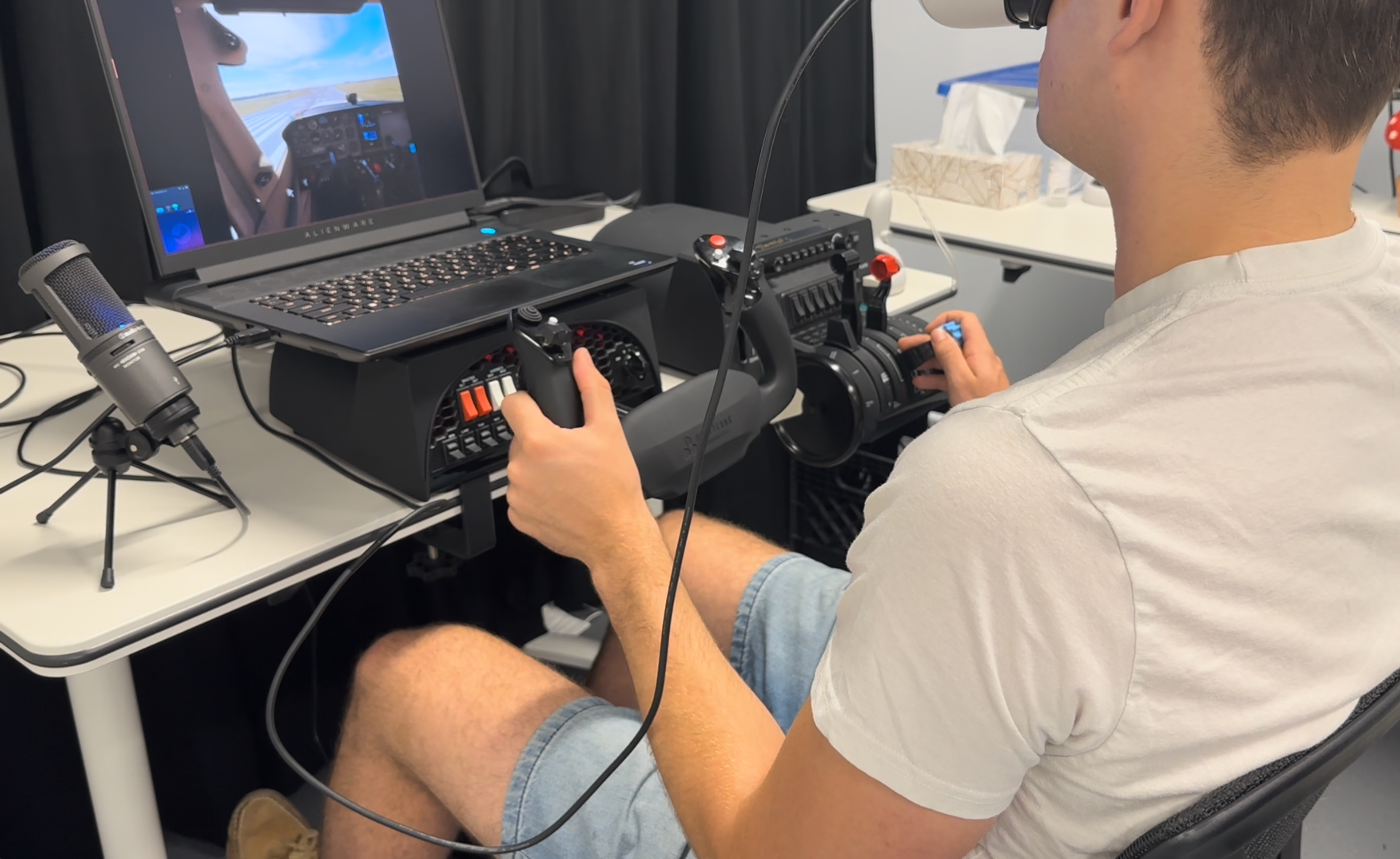}
    \includegraphics[width=0.32\linewidth,trim=0 10 0 10,clip]{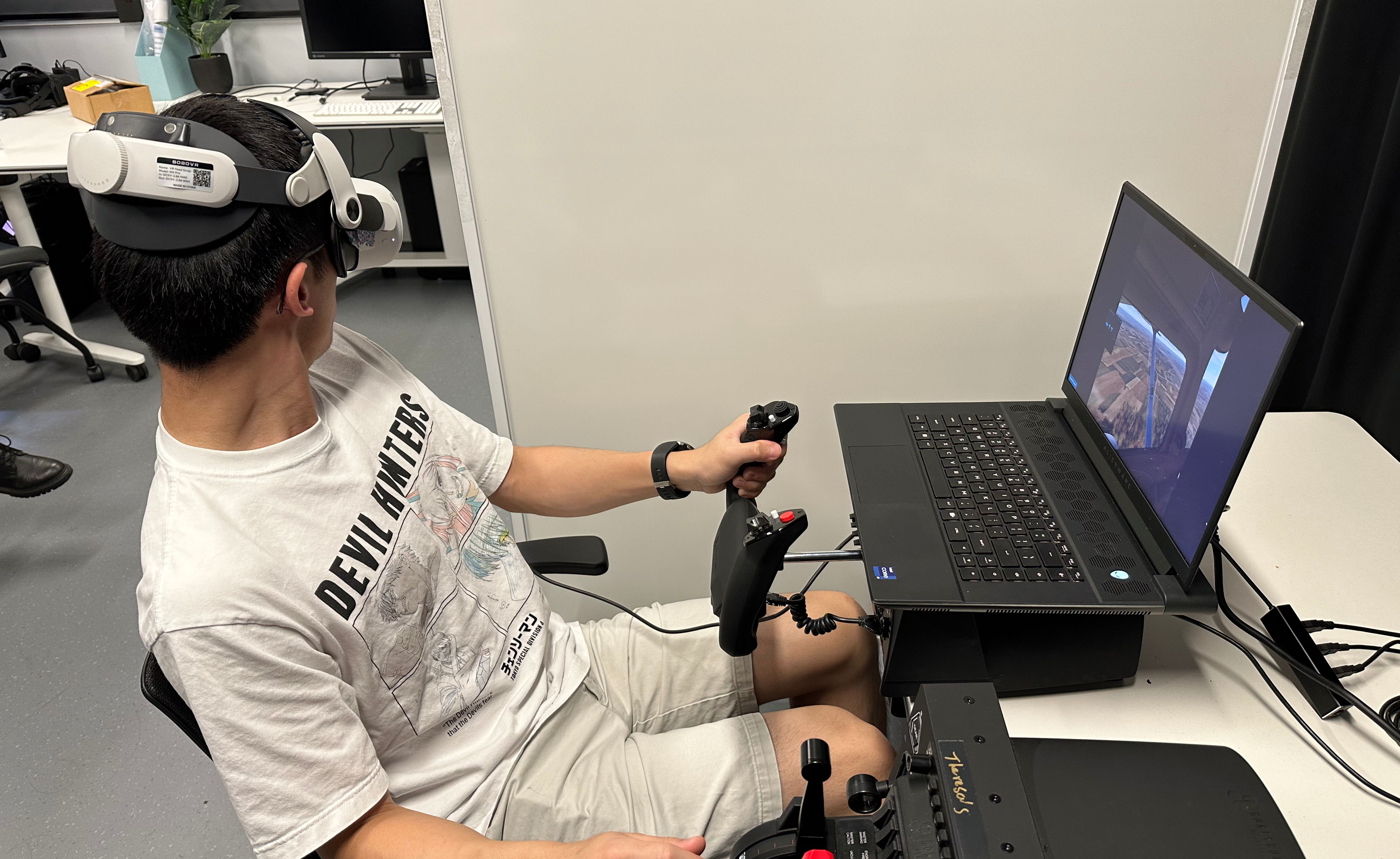}
    \vspace{-3mm}
    \caption{Study sessions with participants. Left: A participant was adjusting the elevator trim, preparing for landing. Center: A participant was landing with the \name{}. Right: A participant was checking the side window for situational awareness.}
    \label{fig:onsite_study}
    \Description{A series of three photographs capturing participants engaged in a simulated flight study. Each participant is using VR equipment and flight control hardware to interact with the simulation. In the left image, a participant adjusts the elevator trim on a detailed flight control setup, preparing for a landing scenario. The center image shows another participant, focusing intently on the monitor while handling the controls during a landing, aided by \name{} displayed on the screen. The right image features a participant checking the virtual side window, enhancing their situational awareness in the simulation. These images collectively depict the immersive and interactive nature of the study, highlighting the use of VR and specific tasks performed during the sessions.}
\end{figure}

\begin{figure}[tb!]
    \centering
    \includegraphics[width=0.45\linewidth,trim=0 10 0 10,clip]{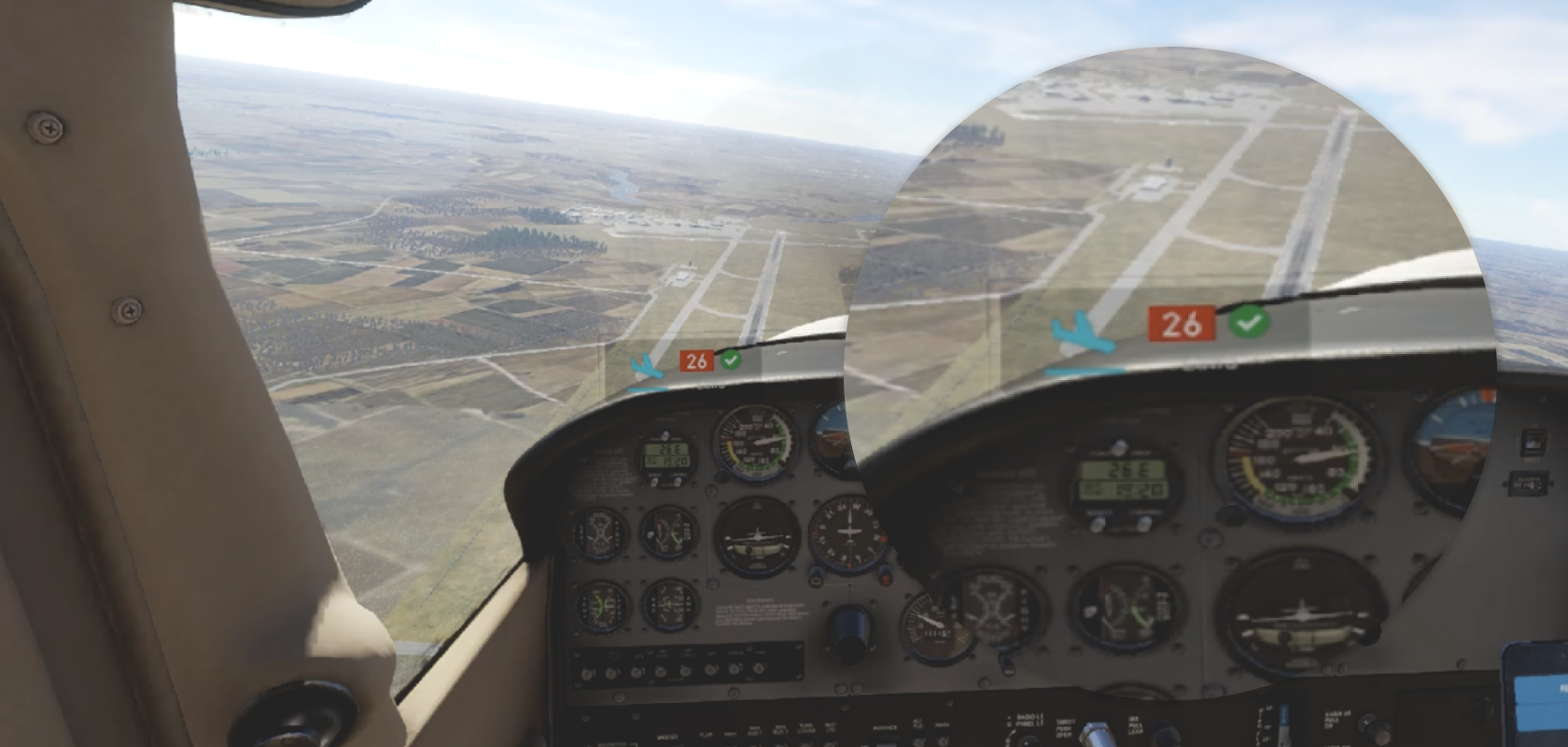}
    \includegraphics[width=0.45\linewidth,trim=0 10 0 10,clip]{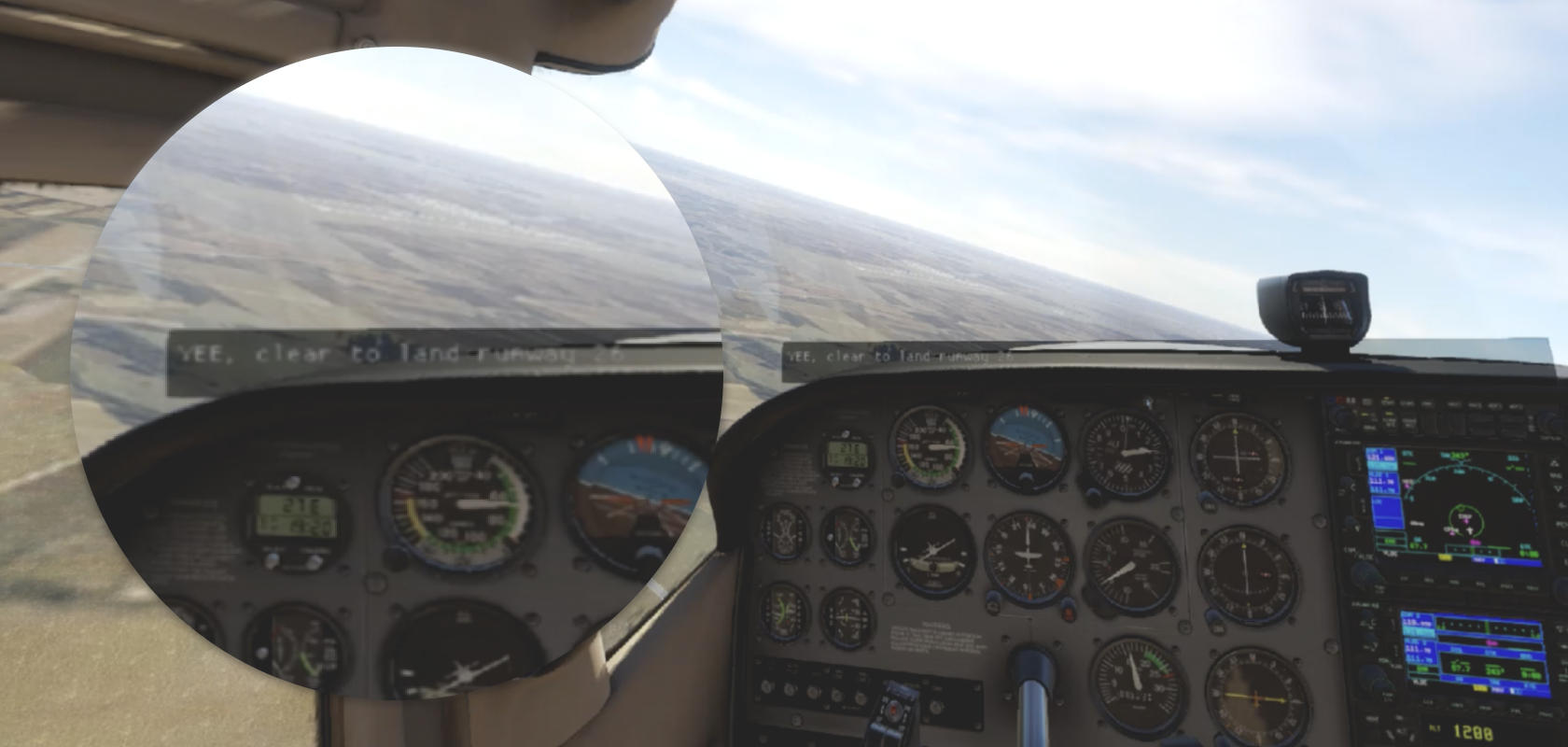}
    \vspace{-3mm}
    \caption{The \name{} (left) and the text-based visual aid (right) are placed above the dashboard in the cockpit's nose area of a Cessna 172 Skyhawk.}
    \label{fig:screenshoot_study}
    \Description{The figure displays two comparative screenshots from a Cessna 172 Skyhawk simulator, demonstrating different types of visual aids. The left image features \name{} overlaying the cockpit's dashboard, with clear and colorful icons such as a green checkmark next to runway number 26, indicating clearance. The right image showcases a text-based visual aid with messages like 'VET, clear to land runway...', placed similarly above the dashboard in the cockpit's nose area. Both images provide a pilot’s view during flight, illustrating how each type of visual aid integrates into the cockpit environment to assist with navigational tasks.}

\end{figure}

\subsection{Conditions}
We established three experimental conditions including the traditional audio-only ATC communication with no visual aids (\textit{\NA}) and those augmented with text-based (\textit{\TA}) and \textit{\name{}}. 
All other variables in the flight simulation, such as weather (clear and calm), landing airport, and aircraft type, were kept constant across the conditions to ensure a fair comparison.
In the \NA{} condition, participants relied solely on verbal instructions from a simulated ATC. 
In the \TA{} condition, visual aids were provided in the form of text that represented the messages, in addition to verbal ATC messages. 
In the \name{} condition, participants received messages supplemented by the designed iconic representation, besides verbal ATC messages.
All visual aids were consistently displayed at the same position, the nose area in the cockpit, directly in the pilot's field of view (\autoref{fig:screenshoot_study}).
The messages delivered in the conditions were standardized to match the typical complexity and length of ATC instructions during approach and landing phases. 

\subsection{Task \& Procedure}
Upon arrival, participants were briefed on the study’s objectives and procedures, and informed consent was obtained.
Participants first engaged in a brief training session to familiarize themselves with the VR setting, controls, simulation software, and overall environment.
Afterward, they experienced the three experimental conditions (\ie, \NA{}, \TA{}, \name{}), presented in a counterbalanced order.
The combinations between the three audio scenarios and the three conditions were randomly determined.
To evaluate the impact of the design, the visual aids (\ie, \TA{} and \name{}) were designed to consistently deliver accurate information.
To encourage pilots to treat audio as the primary channel, the visual aids appeared with a 2-second delay after the audio began and remained visible until the next instruction. 
This timing mimics potential latency in real-world systems (e.g., 300ms--1s)~\cite{ma2023low, liu2020lowlatencysequencetosequencespeechrecognition, yu2021fastemitlowlatencystreamingasr} while maintaining experimental control. 
Although potential implementations may rely on AI transcription or alternative communication pipelines between pilots and controllers, our study focuses on the design and effectiveness of visual representations, not the mechanism by which messages are generated.
Pilots were instructed to read back ATC messages and execute related flight actions while maintaining stable control during realistic flight scenarios.

For \name{}, participants were given supplementary materials to familiarize themselves with the designed icons and visual aids immediately before.
During each condition, participants were instructed to read back ATC messages and execute related flight actions while maintaining stable control during realistic flight scenarios (\autoref{fig:onsite_study}). 
\rev{This setup reflects the multitasking nature of actual cockpit operations, requiring participants to manage all necessary flight tasks (\eg, Fly, Navigate, Communicate) in realistic VFR flight scenarios.}
After completing each condition, participants filled out a questionnaire measuring their workload and user experience. 
Following the completion of all conditions, a semi-structured interview was held to collect qualitative feedback on their experiences with the three different communication methods.
The entire study session lasted about 90 minutes and each participant received \$20 for their time and effort. 
The sessions were video and audio recorded along with data collected on communication performance metrics.
Flight performance data was automatically logged by the simulation system for subsequent analysis.

\subsection{Measures}
To gain a comprehensive understanding of how \name{} supports real-time ATC communication during critical flight phases (particularly its impact on communication effectiveness, cognitive workload, and usability), 
we adopted a mixed-methods approach, combining objective performance metrics, subjective ratings, and qualitative feedback.

\textbf{Objective measures} were collected to reflect on pilot performance, including 1) communication metrics: readback accuracy (how accurately pilots repeated ATC instructions) and readback delay (how promptly they responded), as well as 2) flight performance indicators: airspeed and vertical airspeed, which reflect the stability of aircraft control during flight \cite{morrow1993influence, prinzo2006outcome}.
The scale-independent ``stability'' score was calculated using the coefficient of variation~\cite{stat_cov}, defined as $\frac{\text{SD}(\Delta x)}{\left| \text{Mean}(\Delta x) \right|}$, where $\Delta x_i = x_{i+1} - x_i$, indicating the average stability for airspeed and vertical airspeed across all conditions.
We conducted statistical analysis on these measures to assess the effect of visual aids on both communication and flight performance metrics using appropriate tests.

\textbf{Subjective measures} were gathered via the NASA-TLX questionnaire~\cite{nasa-tlx} using a 7-point Likert scale to assess participants’ perceived cognitive workload in all conditions. 
In addition, participants rated each visual aid (\ie, \name{} and \TA{})'s usability, utility, and clarity on separate 7-point Likert scales.
Similarly, we carried out statistical tests on these measures to compare participants' perception on the three conditions.

\textbf{Qualitative feedback} was obtained base on our semi-structured interviews to understand the pilot’s view of \name{} design in the cockpit, perceived benefits and drawbacks, and suggestions for future improvements.
Open-ended responses were recorded, transcribed, and qualitatively analyzed by two authors using affinity mapping to extract recurring themes and user perceptions.

\subsection{Results \& Discussion}

To better contextualize the results of our design exploration, we present our findings in key themes and discuss their effects: how well \name{} helped pilots enhance ATC communication and perceived, engaged with, and envisioned integrating \name{} into their cockpit, and how these experiences varied across conditions and experience levels, by coordinating both quantitative and qualitative results.
To make the report more concise, detailed results are shown in \autoref{tab:communication_effectiveness_and_performance} 
(for objective measures) as well as in \autoref{tab:nasa} (for subjective measures) in \autoref{apx:quan-analy}.

\begin{figure}[b]
    \includegraphics[width=0.45\textwidth]{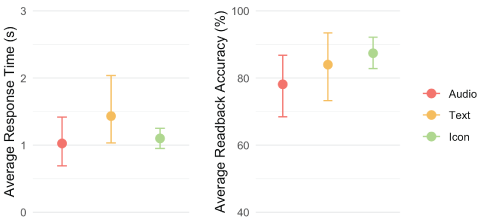}
    \centering
    \vspace{-2mm}
    \caption{Mean readback delay (second; the lower the better) and readback accuracy (the higher the better) for the three conditions. Error bars indicate 95\% confidence intervals.}
    \label{fig:average_response_time_and_reacback_accuracy_95_CI_plot}
    \Description{Left to right: a 95\% confidence interval plot for average ATC message readback accuracy, and a 95\% confidence interval plot for average ATC message response time across three conditions.}
\end{figure}

\textbf{\name{} improved readback accuracy (D1).} 
Our results demonstrate that \name{} significantly reduced miscommunications without compromising communication efficiency or operational performance. 
Notably, participants did not miss any messages in readback under the \name{} condition, whereas three and two messages were missed in the \TA{} and \NA{} conditions, respectively.

\begin{figure}[t]
    \includegraphics[width=0.45\textwidth]{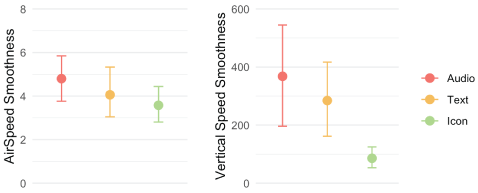}
    \centering
    \vspace{-2mm}
    \caption{Mean stability (the lower the better) of airspeed (knot) and vertical airspeed (feet/second) for the three conditions. Error bars indicate 95\% confidence intervals.}
    \label{fig:flight_performance_95_CI_plot}
    \Description{Left to right: a 95\% confidence interval plot for the stability of airspeed, and a 95\% confidence interval plot for the stability of vertical airspeed across three conditions.}
\end{figure}

Post-hoc Wilcoxon Signed-Rank test showed that \name{} led to significantly higher average readback accuracy than \NA{} ($W=5, p = 0.043$) and also yielded the highest average readback accuracy, followed by \TA{} and \NA{} conditions (\autoref{fig:average_response_time_and_reacback_accuracy_95_CI_plot}).
Importantly, this improvement did not come at the cost of increased cognitive workload or degraded flight performance, where we did not find any significant difference in average readback delay, airspeed and vertical airspeed (\autoref{tab:communication_effectiveness_and_performance}).
As observed, participants maintained the most stable airspeed and vertical airspeed control across the \name{}, \NA{}, and \TA{} conditions (\autoref{fig:flight_performance_95_CI_plot}), indicating that the added visual aids did not interfere with primary flight operations. 

\begin{figure*}[h]
    \includegraphics[width=1\textwidth]{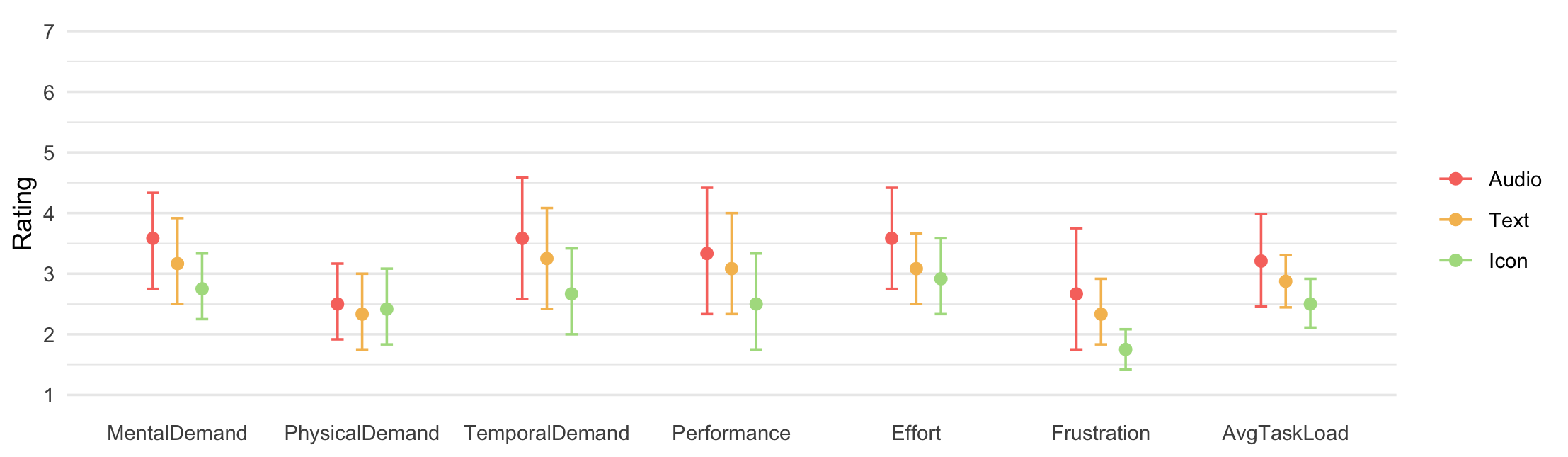}
    \centering
    \caption{NASA-TLX ratings for the three conditions on a 7-point Likert scale (the lower the better). Error bars indicate 95\% confidence intervals.}
    \Description{95\% confidence interval plots for unweighted NASA-TLX ratings on a 7-point Likert scale (the lower the better). }
    \label{fig:nasa}
\end{figure*}

Subjective workload ratings (via NASA-TLX) further supported these findings.
Participants reported the lowest average workload scores under the \name{} condition (\autoref{fig:nasa}), suggesting a subtle but meaningful reduction in perceived cognitive demand.
Together, these results suggest that \name{} enhanced communication accuracy while preserving cognitive and operational performance.

\begin{figure}[b]
    \includegraphics[width=0.45\textwidth]{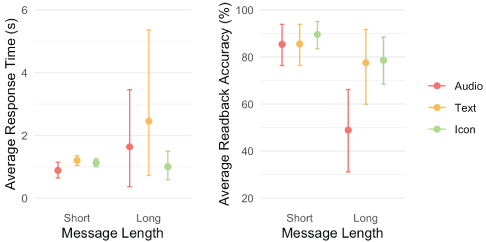}
    \centering
    \vspace{-2mm}
    \caption{Mean readback delay (second; the lower the better) and readback accuracy (the higher the better) for the three conditions, divided by the message length. Error bars indicate 95\% confidence intervals.}
    \label{fig:average_response_time_and_reacback_accuracy_by_length_95_CI_plot}
    \Description{Left: a 95\% confidence interval plot for average ATC message readback accuracy, divided by message length; right: a 95\% confidence interval plot for average ATC message readback daley across three conditions, also divided by message length, with the lower bound of the confidence interval was truncated at zero.}
\end{figure}

\begin{figure*}[]
    \includegraphics[width=1\textwidth]{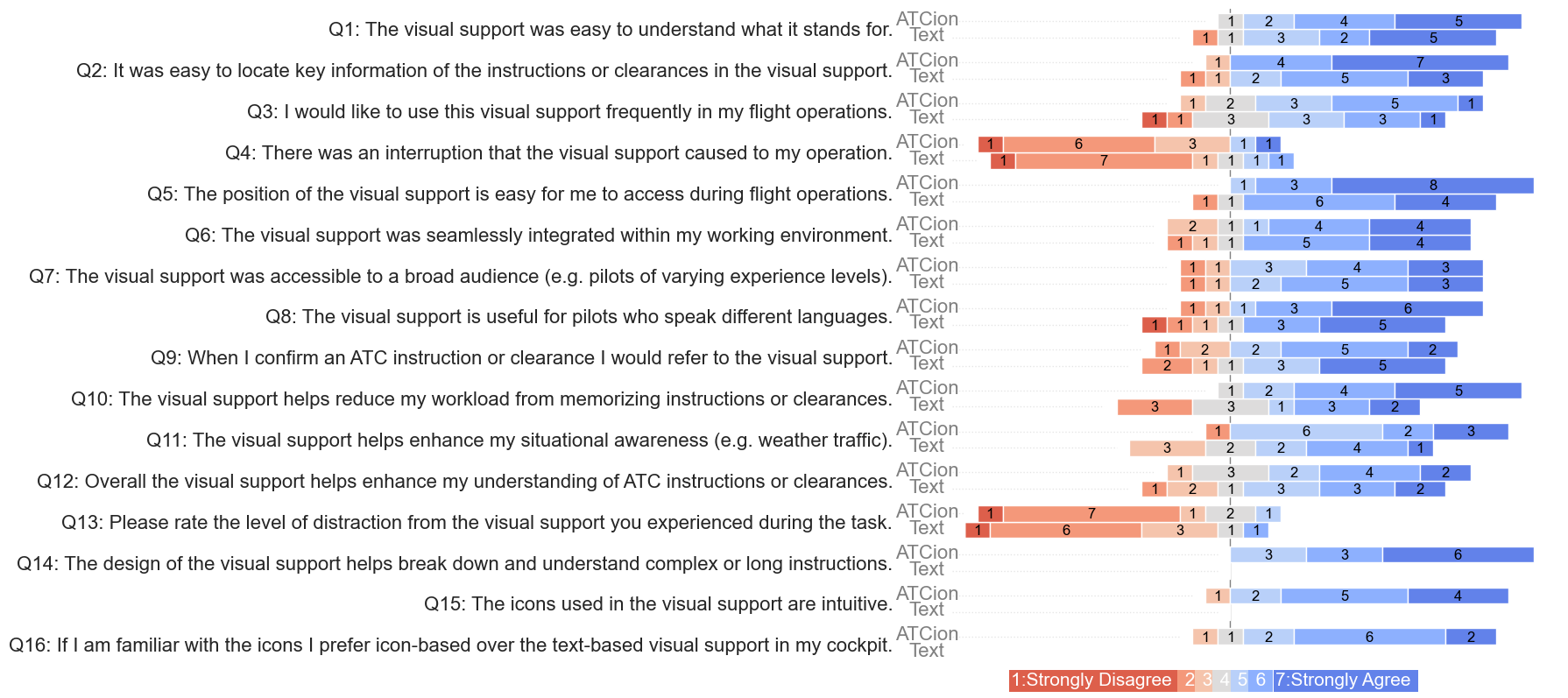}
    \vspace{0mm}
    \caption{Participants' responses to the questionnaire are grouped into four categories: Q1-Q2 access usability, Q3-Q8 explore practical use, Q9-Q13 assess the effectiveness of both visual aids, and Q14-Q16 focus on the specific design features of \name{}. Ratings are provided on a 7-point Likert scale (1: Strongly disagree, 7: Strongly agree), except for Q13 which measures the level of distraction from the visual support (1: No Distraction At All, 7: Very Distractive). While both \name{} and Text-based visual aids provided utility, \name{} appeared to offer potential advantages in several areas, including reducing memorization, supporting information location and confirmation, and improving comprehension. } 
    \centering
    \Description{The figure shows a detailed graphical representation of participant responses to a questionnaire about visual support in the cockpit, organized into four categories: access usability (Q1-Q2), practical use (Q3-Q8), effectiveness of visual aids (Q9-Q13), and design features of \name{} (Q14-Q16). Each question is represented by a horizontal bar graph with segments showing the number of responses from strongly disagree (1) to strongly agree (7). An exception is Q13, which measures the level of distraction from visual support, ranging from no distraction at all (1) to very distractive (7). The colors red and blue distinguish between \name{} and text-based visual supports respectively, illustrating varied levels of agreement or disagreement among participants.}
    \label{fig:ux}
\end{figure*}

\textbf{\name{} reduced memory workload (D1).} 
Beyond overall readback accuracy, our findings highlight how \name{} helped alleviate memory load, particularly during the recall of longer instructions. 
Specifically, average readback accuracy remained the highest in the \name{} condition for both short and long messages (\autoref{fig:average_response_time_and_reacback_accuracy_by_length_95_CI_plot}), those exceeding four terms \cite{cowan2001magical}, a threshold often associated with working memory capacity.
This benefit was especially pronounced for longer ATC messages, where participants typically showed reduced performance due to their memory limitations.
A post-hoc pairwise comparisons with Bonferroni correction showed that, in \NA{}, average readback accuracy dropped significantly from 85.4\% for short messages to 48.9\% for long ones ($t = 4.483, p = .014$), highlighting the challenge of retaining longer instructions without visual support.
In contrast, no significant difference was found between long and short messages in \name{} and \TA{} conditions (\autoref{tab:communication_effectiveness_and_performance}).
These results suggest that both visual aids helped buffer the effects of message length, supporting participants’ ability to retain and recall complex instructions under time pressure.
Notably, a Wilcoxon-Pratt signed-rank test indicated that participants rated \name{} as significantly more effective in reducing memory workload than in \TA{}, as reflected in the responses to Q10 (The visual support helps reduce my workload from memorizing instructions) in \autoref{fig:ux} ($Z = -2.25, p < .05$).

Participants' qualitative feedback reinforced this interpretation.
Several participants described \name{} as intuitive and cognitively efficient. 
P4 noted, \qt{Icons help reduce my workload by allowing me to remember and follow ATC instructions visually}, while P3 highlighted their value in high-demand scenarios \qt{It [\name{}] will be like a tool to help with very long complex instructions.} 
Taken together, these findings indicate that \name{} functioned as a cognitively efficient support tool, especially for longer messages that are more likely to exceed working memory capacity.

\textbf{\name{} was preferred for clarity and usability (D2).}
While most subjective ratings, including overall usability, clarity, and preference, did not show statistical significance, participants consistently trended toward favoring \name{} over the text on preference rating (\autoref{fig:ux}). 
Many found \name{} to be intuitive and easy to use, especially under time pressure, noting that it provided enough structure to be reliable while remaining quick and effortless to interpret.

P4 described the experience plainly \qt{Icons just make it so much easier to look at than reading [text].}
P7 noted, \qt{Even when things got busy, I could still quickly scan the icons and know what to do.}
The chunked layout was particularly appreciated for reducing mental effort. 
P5 shared \qt{I honestly think I would prefer the icons over the text just because they’re broken into chunks, it feels less overwhelming.}
Others emphasized reduced ambiguity and decision fatigue \qt{The icons point you in the right direction without making you think too hard,} said P2.
These comments reinforce that clarity in high-stakes environments is not merely a matter of aesthetics, it is about minimizing cognitive friction when every second counts.

\textbf{Proper placement and semantic alignment shaped the effectiveness of \name{} (D3).} 
Participants highlighted that the utility of \name{} was influenced not only by their presence but also by how and where they were integrated into the interface.
Several emphasized the importance of physical placement within the cockpit environment.
For example, P4 noted that the icons were \qt{right in the middle of two spots—windshield and instruments—you’re going to be looking anyway,} while P7 appreciated that the icons were \qt{within my field of view but not blocking it.}
These comments suggest that unobtrusive but easily glanceable placement contributed to the system’s perceived usability.

However, some participants reported that mismatches between the visual icon sequence and the structure of spoken ATC instructions introduced confusion.
As P6 explained, \qt{The icon order didn’t match the way ATC gives instructions. That made it confusing.}
Such feedback points to the importance of aligning the visual syntax of icon-based representations with the temporal and semantic structure of ATC discourse.
Together, these findings underscore the need to consider both ergonomic integration and linguistic consistency in the design of visual aids for real-time operational contexts.

\textbf{Experience level influenced reliance on \name{}.}
To explore how flight experience shaped participants’ perception with visual aids, we grouped participants using a 100-hour PIC threshold, a common benchmark in aviation training \cite{faa_training}.
Novice pilots appeared to rely more heavily on icons to confirm instructions and reduce memory load.
As P8 noted, \qt{It would be mostly beneficial to pilots who are in their earlier phases of training.}, which was echoed by P6, a flight instructor shared: \qt{A lot of students have a tough time adapting to remembering air traffic control clearances… something like that [\name{}] would definitely help them refresh their memory.} 
Quantitatively, novice pilots achieved their highest average readback accuracy under the \name{} condition ($88.2\%$), compared to \TA{} ($82.4\%$) and \NA{} ($81.0\%$) conditions respectively (\autoref{tab:communication_effectiveness_and_performance}). 

In contrast, experienced pilots reported using visual aids more as a secondary reference.
As P6 (PIC = 350h) stated, \qt{I almost wasn’t even looking at the symbols. I was just focusing on what the ATC was saying.}
He further elaborated that icons and ATC phrases \qt{feel like different languages,} suggesting a cognitive mismatch due to internalized verbal workflows. 
Nonetheless, their average readback accuracy remained consistently high under both \name{} (85.8\%) and \TA{} (87.2\%) conditions, while dropping notably in the \NA{} condition (72.4\%), suggesting that even for experienced pilots, visual aids provided performance benefits when audio cues were insufficient.
Taken together, these findings suggest that visual aids may serve different roles across experience levels: as cognitive scaffolds for novices, and as optional yet performance-enhancing references for experienced pilots, particularly under pressure.



\textbf{Multimodal communication is implied as an integration into future ATC systems.}
Across all data sources, \name{} demonstrated strong potential as part of a multimodal approach to ATC communication.
Our results indicated that participants do not solely rely on auditory ATC instructions but effectively combine them with both visual aids to enhance communication. 
In the multimodal system under investigation, auditory instructions provide the immediate, dynamic information, while the visual layer, comprising both icons and text, serves as a persistent reference that pilots can consult as needed.
P6 remarked, \qt{Compared to radio, it’s like when you’re talking, then it’s gone. Just when the state is up there, you can refer it back,} demonstrating that while the auditory channel delivers real-time messages, the visual channel retains critical information for later confirmation. 
P1 explained that the integrated icon and text display enables rapid scanning: \qt{Each element in the visual aid represents a distinct component of the instruction. This allows me to quickly identify, for instance, my taxi and ground instructions without having to rely solely on memory.}

Furthermore, several participants emphasized that the combined use of visual elements (\ie, icons) and verbal cues helps them manage their workload more efficiently. 
As one pilot noted, the \name{} acts as a \qt{safety net} that supports the fast-paced auditory communication by enabling a quick check-back on prior instructions. 
This synergy between the auditory and visual channels is at the core of the multimodal communication approach: while the audio channel provides dynamic, time-sensitive data, the visual channel, enhanced by the dual presentation of icons and text, offers a complementary, stable reference that reduces memory load and minimizes the risk of miscommunication.
In summary, our results indicate that a multimodal ATC communication system, which integrates both auditory and visual (iconic and textual) elements, can aid pilots in high workload scenarios. 
This dual-channel approach leverages the strengths of each modality: immediate responsiveness via audio and confirmatory recall through visual supports.



\section{Towards future in-cockpit visual aid design} \label{sec:dicussion}
Our study results confirm the effectiveness of \name{} in ATC communication on various aspects, and provide insights into the utility of visual aids.
\rev{We position this work as early-stage research that explores design opportunities and informs future systems in cock-pit visual aids design.
While we envision the potential deployment in the future, we acknowledge that real-world adoption will ultimately require alignment with certification standards (\eg, FAA Part 23~\cite{cfr23, ac231311}, EASA CS‑23~\cite{cs23}).}
In this section, we discuss the implications of our study on future research and practical applications, as well as the design components where improvements could be made. 

\subsection{Design Implications}
\textbf{Optimizing the Use of Text and Icons in ATC Communication.}
Building on previous research that predominantly focused on text-based systems~\cite{kerns1991data,prinzo2001data}, our study is among the first to comprehensively examine the efficacy of icons in real-time ATC communication. 
The findings reveal that icon-based visual aids were generally preferred for delivering ATC instructions, offering improved readback accuracy and reduced memory workload compared to text-only presentations.

However, text remains essential for conveying atypical, nuanced, or less standardized directives---highlighting the limitations of a purely icon-based system. 
Our results point toward the value of a hybrid approach that leverages icons for routine or structured communications while reserving text for more complex or exception-based cases. 
Such a strategy could help balance operational efficiency with communication accuracy, especially under varying workload conditions.
This hybrid approach is particularly relevant in the context of emerging aviation systems such as the Next Generation Air Transportation System (NextGen)~\cite{navarro1999datalink, hoogeboom2004does}, where datalink and advanced communication technologies are reshaping pilot-controller interactions. 
Optimizing the integration of text and icons within these systems could not only improve communication clarity but also enhance safety, reliability, and overall system performance. 
Future work should systematically investigate how to best implement hybrid icon-text strategies across varying instruction types, operational contexts, and workload conditions, particularly within advanced systems like NextGen.

\textbf{Designing for Attention, Memory, and Action in Visual ATC Aids in Cockpits.}
While \name{} offers a promising step toward integrating multimodal support in the cockpit, participants’ responses point to several broader design and interaction considerations that can inform future work in aviation and safety-critical domains.
One insight concerns the perceptual salience of information under time pressure. 
Participants indicated that rapid recognition of message urgency or priority could be supported through visual encoding techniques such as color or spatial grouping. 
Prior research in perceptual design and interface salience suggests that such techniques can improve information triage and reduce reaction time in high-stakes scenarios~\cite{Macdonald1988Evaluating}. 
This calls for further exploration of how visual encodings may modulate attentional shifts in-cockpit environments, especially in combination with existing auditory channels.

Another recurrent theme was the temporal dimension of communication. 
Participants expressed a desire to access prior transmissions, either their own or those directed to nearby aircraft, during low-workload phases. 
This finding suggests that persistent visual traces of communication may serve as cognitive scaffolding, supporting retrospective sensemaking and collaborative awareness~\cite{Narborough-Hall1985Recommendations}. 
Such persistence aligns with HCI work on shared displays and asynchronous collaboration, where memory aids help bridge temporal gaps in distributed coordination.

Participants also emphasized the importance of information relevance and density, highlighting the need for dynamic information management. 
Interfaces that continually update to foreground the most contextually relevant messages—while suppressing or fading less critical ones—could help reduce clutter and mitigate visual overload. 
This reinforces prior findings that selective redundancy, rather than cumulative layering, is key to maintaining usability in information-rich environments~\cite{Luder1984Redundant}.

Collectively, these observations point to broader design implications beyond this particular implementation. 
Designers of cockpit systems, and more generally, of multimodal interfaces in high-risk domains, must carefully balance augmentation with cognitive economy. Visual aids should not merely replicate verbal content but transform it into formats optimized for perceptual efficiency and situational recall. 
Furthermore, the heterogeneity in participant preferences underscores the need for configurable or adaptive interfaces that respond to expertise, context, and individual strategies for workload management.
Future research should investigate how such adaptive systems perform across different levels of pilot expertise and phases of flight.


\textbf{Implementing AI-Based Transcription as a Potential Approach.}
Although \name{} currently operates with scripted inputs, its demonstrated benefits---such as improved communication accuracy and reduced memory load---suggest strong potential for real-world application. One promising direction for enabling deployment at scale is the use of AI-based speech recognition to automatically transcribe ATC communications and generate corresponding visual aids.
Recent advances in transcription accuracy, including real-time call-sign detection~\cite{spille2018comparing, kasttet2023toward}, indicate growing feasibility for such automation in operational environments. 
This could allow systems like \name{} to dynamically visualize incoming instructions, reducing pilot workload while preserving communication clarity.
This dynamic visualization not only streamlines the interpretation process but also lays the groundwork for more proactive and context-aware communication support.

Building on this foundation, future implementations might extend these capabilities even further by transmitting visual aids directly from ATC to the cockpit. 
Such an extension could provide controllers with the ability to tailor visual messages to current conditions, thereby enhancing shared situational awareness and coordination between pilots and controllers. 
As these technologies continue to mature, integrating AI-driven transcription and visualization could facilitate the broader adoption of multimodal communication tools like \name{}, particularly within evolving frameworks such as NextGen.

\subsection{Limitations and Future Work}
While this study highlights the potential of visual aids, particularly icons, in supporting ATC communication, several limitations remain that warrant further exploration.
First, due to challenges in recruiting participants with varied experience levels, the study involved a relatively small and homogeneous sample, limiting the generalizability of our findings. 
Future research should involve a larger, more diverse participant pool with controlled experience levels to better understand how experience influences the perceived utility and effectiveness of visual aids.
Second, the study focused on approach and landing phases, excluding routine operations like cruising and critical edge cases such as emergencies. 
Although cruise involves lower workload, display strategies may differ, and visualizing less urgent messages could still support situational awareness. 
Edge cases, though infrequent, are vital for system resilience and safety. 
Exploring these scenarios in future work can help ensure robust performance across various operational contexts.
Third, our evaluation was conducted in a short-term, high-fidelity VR-based setup that limited participants’ familiarity with the icons and did not fully replicate real-world conditions.
Longitudinal studies are needed to examine how trust, reliance, and behavior evolve with continued use of visual augmentation. 
Such research would enhance both design and our understanding of how interface modalities shape cognition and action in safety-critical environments.
Finally, while our evaluation covered effectiveness, flight performance, workload, and situational awareness, future work could incorporate additional metrics. 
These could include the frequency and type of miscommunications, the cognitive demands of processing visual aids, and awareness across different flight phases. 
Such measures would provide deeper insight into trade-offs between communication speed, accuracy, and efficiency.

\section{Conclusion}
We proposed an icon-based visual aid approach to support in-cockpit ATC communication and explored its integration into real-time pilot workflows.
Through an iterative, user-centered design process with domain experts, we developed \name{} and established practical design guidelines for visualizing ATC messages.
Through a user study in VR-simulated flight scenarios, we have demonstrated that \name{} improves communication effectiveness, lowers cognitive strain, and supports stable flight performance during critical phases.
These findings highlight the potential of icon-based aids to enhance situational awareness and reduce miscommunication in high-stakes aviation environments, laying the groundwork for future research on visual support in real-time communication systems.

\begin{acks}
This work is supported in part by 
the Natural Sciences and Engineering Research Council of Canada (NSERC) Discovery Grant \#RGPIN-2020-03966, 
the Waterloo Institute for Sustainable Aeronautics (WISA) FedDev RESEARCH-FOR-IMPACT (RFI) Program, and 
NSF grant IIS-2441310.
We thank all participating pilots and ATC professionals for their time and valuable insights. 
We are also grateful to Cara Li for her support and collaboration throughout the project. 
Finally, we thank the reviewers for their thoughtful feedback, which significantly improved the quality of this paper.
\end{acks}

\bibliographystyle{ACM-Reference-Format}
\bibliography{references.bib}

\clearpage
\appendix
\section{Types and Characteristics of ATC Messages} \label{apx:atc-messages}
The seven main types of ACT messages are as follows. \autoref{tab:ATC_characteristics} further highlights their basic characteristics including timeliness (\eg, how quickly the message becomes outdated), action timing (\eg, when the message should be acted upon), relevance to flight phases (\eg, during which phase the message is most likely received), and impact on pilot workload (\eg, how busy the pilot is when receiving the message). 

\textbf{Instructions}, such as altitude changes or maneuver directives, are action-oriented messages that require the pilot's immediate attention. 
The experts agreed that these messages should be presented in a way that facilitates quick comprehension and execution. 
Visual aids for instructions should be prominently positioned either in fixed locations within the cockpit or the pilot's field of view (FOV) to ensure clear, immediate access.
This allows pilots to reference critical actions without diverting focus from their primary tasks.

\textbf{Clearances}, which authorize flight paths, altitudes, or other critical parameters, were deemed important but not as urgent as emergency messages. 
The questionnaire results indicated that while clearances need to be acted upon promptly, they do not necessarily require the same immediate visibility. 
The experts preferred that clearance information be displayed in fixed locations within the cockpit, allowing easy reference without overwhelming.

\textbf{Requests}, such as inquiries for specific actions or information, were regarded as moderately important. 
The experts suggested that while these messages should be accessible, they do not need to be as prominently displayed as emergency information or instructions. 
Requests were best suited for placement in a secondary location within the cockpit, where they could be quickly referenced when needed.

\textbf{Acknowledgments}, which confirm the receipt and understanding of a message, were viewed as lower priority in terms of visual aids. 
The experts emphasized that these messages often require minimal cognitive load and do not need to remain visible for extended periods.
Acknowledgments can be displayed in brief and disappear once confirmed.

\textbf{General Information}, such as traffic updates or weather reports, was considered useful for maintaining situational awareness but not time-critical. 
The experts preferred that this type of information be presented in a less intrusive area, such as the periphery of the cockpit or on external displays. 
This allows pilots to reference the information as needed without cluttering their primary workspace.

\textbf{Advisories}, which alert pilots to potential hazards or changes in flight conditions, were deemed important but not as urgent as emergency messages. 
The experts suggested that advisories should be displayed in fixed locations in the cockpit or attached to the physical world, allowing them to be easily checked. 
For example, traffic advisories and airport navigation data should be attached to physical elements for quick spotting when necessary.

\textbf{Emergency}, such as those relating to critical situations (e.g., go-arounds or warnings), were universally agreed to be the most time-sensitive and important. 
The experts emphasized the need for these messages to be displayed prominently within the FOV, ensuring immediate visibility and fast response. 
The ability to quickly recognize and act on emergency communications was a key design consideration, and the experts strongly preferred that such information remain visible until fully addressed.

\begin{table}[tb]
\caption{Mode scores across five dimensions for an understanding of ATC messages.
1) Timeliness (TL.): The message/instruction should be 1:Never Outdated - 5:Outdated Immediately;
2) Action Timing (AT.): The message/instruction should be 1:No Need To Apply - 5:Apply Immediately;
3) Flight Phase (FP.): During \_\_  phase, I most likely receive receive this message. 1:On-ground, 2:Takeoff, 3:En-route, 4:Approach, 5:Landing, and 6:Any Above Phase; 
4) Mental Workload Level (WL.): How busy you are on your primary task in the general case when you receive the message, that you could pay less attention to the communication? 1:Relax - 5:Busy}
\label{tab:ATC_characteristics}
\footnotesize
\centering
\setlength{\tabcolsep}{3pt}
\begin{tabular}{p{1.9cm}p{2.5cm}cccc}
\toprule
\textbf{Type} & \textbf{Message} & \textbf{TL.} & \textbf{AT.} & \textbf{FP.} & \textbf{WL.} \\
\midrule
\multirow{14}{*}{Instruction} 
& Squawk code & 2 & 5 & 1 & 1 \\
& Holding instructions & 3 & 5 & 1 & 2 \\
& Runway lineup & 2 & 4 & 1 & 2 \\
& Taxi instructions & 4 & 4 & 1 & 3 \\
& Altitude instructions & 3 & 5 & 4 & 3 \\
& Speed instructions & 3 & 5 & 4 & 3 \\
& Vectoring & 3 & 5 & 3 & 3 \\
& Turn instructions & 3 & 5 & 4 & 4 \\
& Route change & 4 & 2 & 3 & 3 \\
& Direct to waypoint & 4 & 4 & 3 & 3 \\
& Approach instructions & 3 & 5 & 4 & 4 \\
& Holding pattern & 3 & 2 & 4 & 4 \\
& Overshoot & 5 & 5 & 4 & 5 \\
& Freq. change & 2 & 4 & 2 & 3 \\
\midrule
\multirow{3}{*}{Request} 
& Status report & 4 & 4 & 3 & 3 \\
& Squawk ident & 5 & 5 & 4 & 3 \\
& Intention report & 5 & 5 & 3 & 3 \\
\midrule
\multirow{5}{*}{Clearance} 
& Taxi clearance & 3 & 4 & 1 & 3 \\
& Takeoff clearance & 3 & 5 & 1 & 2 \\
& Landing clearance & 3 & 2 & 4 & 5 \\
& Route clearance & 3 & 2 & 3 & 3 \\
& Pushback clearance & 3 & 4 & 1 & 2 \\
\midrule
\multirow{3}{*}{Gen. Info} 
& Call-sign & 1 & 1 & 1 & 3 \\
& Traffic info & 2 & 1 & 3 & 3 \\
& Weather info & 2 & 1 & 4 & 3 \\
\midrule
\multirow{4}{*}{Advisory} 
& Weather adv. & 2 & 2 & 3 & 3 \\
& Safety alert & 3 & 5 & 4 & 4 \\
& Approach adv. & 3 & 5 & 4 & 4 \\
& Traffic adv. & 2 & 2 & 4 & 4 \\
\midrule
\multirow{2}{*}{Emergency} 
& Emerg. info & 2 & 5 & 1 & 3 \\
& Emerg. instr. & 5 & 5 & 1 & 5 \\
\midrule
\multirow{4}{*}{Acknowledgments} 
& Roger & 5 & 1 & 1 & 2 \\
& Unable & 4 & 1 & 3 & 3 \\
& Affirmative & 5 & 1 & 3 & 3 \\
& Approved & 5 & 2 & 1 & 2 \\
\bottomrule
\end{tabular}
\vspace{-1mm}
\Description{The table summarizes mode scores for understanding ATC messages across five dimensions: Timeliness, Action Timing, Flight Phase, Mental Workload, and Message Type which includes Instructions, Requests, Clearances, General Information, Advisories, Emergency, and Acknowledgment. Each row lists a specific ATC message alongside scores that measure its urgency, the immediate need for action, the typical flight phase during which it is received, and the mental workload it imposes on the pilot. The ratings help illustrate how messages differ in their immediacy and the attention they demand, with some messages like Emergency Instructions requiring immediate attention and high mental engagement, contrasted with more routine messages like Call-sign information which are less urgent and demand less mental effort.}
\end{table}

\begin{table*}[!bt]
\caption{The listed terms of ATC messages are commonly used in ATC communication, such as terms in actions or concepts, directions and units. The checked terms are critical information for visualization, voted by experts during the co-design.}
\centering
\small
\label{tab:action}
\begin{minipage}[t]{0.45\linewidth} 
\small
\begin{tabular}{l|r|l}
Type & Term & Visualization\\
\toprule
\multirow{36}{*}{\makecell[l]{Actions/\\Concepts}} & Taxi (via) & \cmark \\
                                    &  Cross & \cmark \\
                                    &  Hold short & \cmark \\
                                    &  Line up                         & \cmark  \\
                                    &  Cleared (to)                    & \cmark  \\
                                    &  Maintain                        &         \\
                                    &  Continue                        &         \\
                                    &  Climb (to)                      & \cmark  \\
                                    &  Descend (to)                    & \cmark  \\
                                    &  Heading                         & \cmark  \\
                                    &  Turn                            & \cmark  \\
                                    &  Extend                          &         \\
                                    &  Cross                           & \cmark  \\
                                    &  Contact                         &         \\
                                    &  Wind check                      &         \\
                                    &  Exit                            & \cmark  \\
                                    &  Above/below                     & \cmark  \\
                                    &  Not above/ below                & \cmark  \\
                                    &  Go-around                       & \cmark  \\
                                    &  Report                          & \cmark  \\
                                    &  Orbit                           &   \\
                                    &  Follow                          &   \\
                                    &  Direct (to)                     &   \\
                                    &  Reduce (speed)                  & \cmark  \\
                                    &  Advise                          &   \\
                                    &  Squawk                          &   \\
                                    &  Stop                           & \cmark  \\
                                    &  Cancelled                       & \cmark  \\
                                    &  Number                          & \cmark  \\
                                    &  Change (frequency)              &   \\
                                    &  Stand-by                        &   \\
                                    &  Enter                          & \cmark  \\
                                    &  Wait                           & \cmark  \\
                                    &  Confirm                         &   \\
                                    &  Reset                           &   \\
                                    &  Report / Say                    &   \\
                                    \cline{1-2}
\multirow{8}{*}{Directions}        &  North                           & \cmark  \\
                                   &  North-East                      & \cmark  \\
                                   &  East                            & \cmark  \\
                                   &  South-East                      & \cmark  \\
                                   &  South                           & \cmark  \\
                                   &  South-West                      & \cmark  \\
                                   &  West                            & \cmark  \\
                                   &  North-West                      & \cmark  \\
 \bottomrule
\end{tabular}
\end{minipage}
\quad
\begin{minipage}[t]{0.45\linewidth} 
\centering
\small
\begin{tabular}{l|r|l}
Type & Term & Visualization \\
\toprule
\multirow{24}{*}{\makecell[l]{Action/\\Concept\\variables}} &  Takeoff  & \cmark  \\
                                   &  Land                            & \cmark  \\
                                   &  Departure                       & \cmark  \\
                                   &  Runway                          & \cmark  \\
                                   &  Right                           & \cmark  \\
                                   &  Left                            & \cmark  \\
                                   &  Taxiway (A-Z)                   & \cmark  \\
                                   &  Speed                           & \cmark  \\
                                   &  Altimeter                       & \cmark  \\
                                   &  Flight level                    & \cmark  \\
                                   &  Frequency                       &  \\
                                   &  Wind                            &  \\
                                   &  Left/right (on/hand)            & \cmark  \\
                                   &  Upwind                          & \cmark  \\
                                   &  Crosswind                       & \cmark  \\
                                   &  Downwind                        & \cmark  \\
                                   &  Base                            & \cmark  \\
                                   &  Final                           & \cmark  \\
                                   &  Circuit                         & \cmark  \\
                                   &  Delay                           & \cmark  \\
                                   &  Traffic                         & \cmark  \\
                                   &  Touch-and-go                    & \cmark  \\
                                   &  Stop-and-go                     & \cmark  \\
                                   &  Flight plan                     & \\
                                   \cline{1-2}
\multirow{5}{*}{numbers}           &  (runway) x-x                    & \cmark  \\
                                   &  (heading) x-x-x                 & \cmark  \\
                                   &  (frequency) x-x-x-decimal-x-x-x & \cmark  \\
                                   &  (wind) x                        & \cmark  \\
                                   &  x-thousand-x- hundred           & \cmark  \\
                                   &  (flight level) x-x-x            & \cmark  \\
                                   &  (taxiway) A-Z                   & \cmark  \\
                                   \cline{1-2}
\multirow{6}{*}{Units}             &  Feet                            &  \\
                                   &  Degree                          &  \\
                                   &  Knots                           &  \\
                                   &  O’clock                         &  \\
                                   &  Miles                           &  \\
                                   &  Zulu                            &  \\
                                   \cline{1-2}
\multirow{7}{*}{Other}             &  When able                       & \cmark  \\
                                   &  Terminated                      &  \\
                                   &  Approved                        &  \\
                                   &  Identified                      &  \\
                                   &  Immediate                       & \cmark  \\ 
                                   &  &   \\ 
                                   &  &   \\ 
\bottomrule
\end{tabular}
\Description{The table lists commonly used terms in ATC communication, categorized into Actions/Concepts, Directions, Action/Concept Variables, Numbers, Units, and Other terms. Each term is assessed for its criticality in visualization, based on expert votes during the co-design phase. Terms marked with a check (✓) are deemed critical for visualization. The table is divided into two sections: the left section includes Actions/Concepts such as "Taxi," "Cross," and "Hold short," and Directions like "North" and "East." The right section includes variables related to actions and concepts like "Takeoff" and "Land," various numerical expressions for runway numbers, headings, and frequency, and units such as "Feet" and "Degrees." Critical terms such as "Departure," "Runway," and directional indicators are prioritized for visual representation to aid in navigation and compliance with ATC instructions.}

\end{minipage}
\end{table*}
\clearpage

\onecolumn
\section{Measures Results} 
\label{apx:quan-analy}
\vspace{1mm}

\begin{table*}[h]
\small
\centering
\begin{tabular}{lrr|rr|rr}
 \toprule
& \multicolumn{2}{c|}{\textbf{Audio}} & \multicolumn{2}{c|}{\textbf{Text}}  & \multicolumn{2}{c}{\textbf{\name{}}} \\ 
 \textbf{} & \textbf{Mean} & \textbf{CIs} & \textbf{Mean}   & \textbf{CIs} &  \textbf{Mean} & \textbf{CIs} \\ 
 \midrule
Mental Demand     & 3.58  & [2.75 4.33]& 3.17  &[2.5, 3.92] & 2.75  & [2.25, 3.33]\\ 
Physical Demand   & 2.5  & [1.92 3.17] & 2.33  &[1.75, 3] & 2.42  & [1.83, 3.08]\\ 
Temporal Demand   & 3.58 & [2.58, 4.58] & 3.25  &[2.42, 4.08] & 2.67  & [2, 3.42]\\ 
Performance       & 3.33 & [2.33 4.42] & 3.08  &[2.33, 4] & 2.5  & [1.75, 3.33]\\ 
Effort            & 3.58 & [2.75 4.42] & 3.08 &[2.5, 3.67] & 2.92  & [2.33, 3.58]\\ 
Frustration       & 2.67 & [1.75 3.75] & 2.33  &[1.83, 2.92] & 1.75 & [1.42, 2.08]\\ 
Avg. Task Load    & 3.21&  [2.46 3.99] & 2.88  &[2.44 3.31] & 2.5 & [2.11, 2.92]\\ 
\bottomrule
\end{tabular}
\vspace{2mm}
\caption{NASA-TLX ratings for the three conditions on a 7-point Likert scale (the lower the better). Error bars indicate 95\% confidence intervals.
}
\label{tab:nasa}
\end{table*}

\begin{table*}[h]
\small
\vspace{1mm}
\centering
\caption{Comparison of key performance metrics across three conditions with mean and 95\% confidence intervals. Metrics include average readback delay (lower is better), average readback accuracy (higher is better), airspeed (lower is better), and vertical airspeed (lower is 
``Short'' and ``Long'' refer to ATC messages with four or fewer terms and more than four terms, respectively.
``Novice'' are defined as those participants with at least 100 hours of PIC (Pilot-in-Command) experience, while ``Experienced'' have over 100 hours.
Readback delay was calculated as the duration between the end of a given instruction and the start of the participant's readback, providing insights into cognitive processing and the ease of formulating a readback.
Readback accuracy was measured by evaluating the percentage of correctly recalled key terms from the ATC messages with respect to the number of total required key terms (a total of 21 terms for each condition).
}
\vspace{-3mm}
\begin{tabular}{lrr|rr|rr|rr}
 \toprule
& \multicolumn{2}{c|}{\textbf{Audio}} & \multicolumn{2}{c|}{\textbf{Text}}  & \multicolumn{2}{c}{\textbf{\name{}}} \\ 
 \textbf{} & \textbf{Mean} & \textbf{CIs} & \textbf{Mean}   & \textbf{CIs} &  \textbf{Mean} & \textbf{CIs} & \textbf{Statistic} & \textbf{p-value} \\ 
 \midrule
Readback delay               & 1.02 & [0.692, 1.41] & 1.44 &[1.02, 2.03] & 1.1 & [0.95, 1.25] & $\chi^2$ = 3.304 &.192\\ 
Readback delay  (Short)       & 0.882 & [0.632, 1.15]& 1.20 & [1.04, 1.35]& 1.12 & [1, 1.27] & & \\ 
Readback delay  (Long)        & 1.64 & [0.364, 3.45] & 2.45 & [0.727, 5.36]& 1 & [0.583, 1.5] &&\\
Readback delay  (Novice)        & 0.875 & [0.55, 1.3] & 1.22 & [1.03, 1.42] & 1.1 & [0.9, 1.27]& & \\
Readback delay  (Experienced)   & 1.32 & [0.675, 2.1]  & 1.85 & [0.8, 3.55] & 1.1 & [0.9, 1.3] & & \\
\cline{1-1}
Avg. readback accuracy           & 78.1\% & [0.685, 0.869] & 84.0\% & [0.733, 0.934] & 87.4\% & [0.826, 0.919] & $\chi^2$ = 9.6 & < 0.01
\\
Avg. readback accuracy (Short)   & 85.4\% & [0.76, 0.936] & 85.6\% & [0.764, 0.941] & 89.6\% & [0.837, 0.951] &  & \\
Avg. readback accuracy (Long)    & 48.9\% & [0.316, 0.662] & 77.5\% & [0.597, 0.917] & 78.6\% & [0.684, 0.883] &  &\\
Avg. readback accuracy  (Novice)       & 81.0\% & [0.704, 0.900] & 82.4\% & [0.687, 0.946] & 88.2\% & [0.817, 0.947] &  & \\
Avg. readback accuracy  (Experienced)  & 72.4\% & [0.558,0.889] & 87.2\% & [0.721, 1]& 85.8\% & [0.904, 0.904] & & \\
\midrule
Airspeed                    & 4.80 & [3.78, 5.87] & 4.06 &  [3.03, 5.25] & 3.58 & [2.78, 4.43] & F =1.28   &.298\\ 
Vertical airspeed           & 368 & [200, 547] & 285 & [159, 414] & 85.8 & [53.7, 124] & F = 1.489  &.247\\
\bottomrule
\label{tab:communication_effectiveness_and_performance}
\end{tabular}

\begin{tabular}{lrr|rr|rr}
 \toprule
& \multicolumn{2}{c|}{\textbf{Audio}} & \multicolumn{2}{c|}{\textbf{Text}}  & \multicolumn{2}{c}{\textbf{\name{}}} \\ 
 \textbf{} & \textbf{Mean} & \textbf{CIs} & \textbf{Mean}   & \textbf{CIs} &  \textbf{Mean} & \textbf{CIs}\\ 
 \midrule
Avg. readback accuracy (Short)   & 85.4\% & [0.76, 0.936] & 85.6\% & [0.764, 0.941] & 89.6\% & [0.837, 0.951] \\
Avg. readback accuracy (Long)    & 48.9\% & [0.316, 0.662] & 77.5\% & [0.597, 0.917] & 78.6\% & [0.684, 0.883]\\
\midrule
Statistic & t.ratio= 4.483 & p = 0.014 &  t.ratio = 1.577 & p = 1 &  t.ratio = 1.633 & p = 1 \\
\bottomrule
\end{tabular}
\end{table*}

\end{document}